\documentclass[aps,pre,onecolumn,showpacs,showkeys,a4paper]{revtex4-1}

\usepackage[centertags]{amsmath}
\usepackage{amsfonts}
\usepackage{amssymb}
\usepackage{amsthm}
\usepackage{newlfont}
\usepackage{stmaryrd}
\usepackage{mathrsfs}
\usepackage{mathtools}
\usepackage{euscript}
\usepackage{graphicx}
\usepackage{enumerate}
\usepackage{tikz}
\usepackage{pgf}
\usepackage{mathtools}
\usetikzlibrary{positioning,fit,calc}
\usetikzlibrary{arrows,automata}
\usepackage{wrapfig}
\usepackage{subfigure}
\usepackage{amscd}
\usepackage{hyperref}

\theoremstyle{plain}

\theoremstyle{definition}

\theoremstyle{remark}

\newcommand{\opunit}{\text{1}\kern-0.22em\text{l}}

\DeclareMathAlphabet{\mathpzc}{OT1}{pzc}{m}{it}

\begin{document}

\title{Large deviations and dynamical phase transitions in stochastic chemical networks}
\author{Alexandre Lazarescu$^{(1)}$, Tommaso Cossetto$^{(2)}$, Gianmaria Falasco$^{(2)}$ and Massimiliano Esposito$^{(2)}$}
\affiliation{(1) Centre de Physique Théorique, CNRS UMR 7644, \'Ecole Polytechnique, F-91128 Palaiseau Cedex, France\\
(2) Complex Systems and Statistical Mechanics, Physics and Material Science Research Unit, University of Luxembourg, L-1511 Luxembourg, Luxembourg}
\pacs{}
\keywords{}
\begin{abstract}
Chemical reaction networks offer a natural nonlinear generalisation of linear Markov jump processes on a finite state-space. In this paper, we analyse the dynamical large deviations of such models, starting from their microscopic version, the chemical master equation. By taking a large-volume limit, we show that those systems can be described by a path integral formalism over a Lagrangian functional of concentrations and chemical fluxes. This Lagrangian is dual to a Hamiltonian, whose trajectories correspond to the most likely evolution of the system given its boundary conditions. 

The same can be done for a system biased on time-averaged concentrations and currents, yielding a biased Hamiltonian whose trajectories are optimal paths conditioned on those observables. The appropriate boundary conditions turn out to be mixed, so that, in the long time limit, those trajectories converge to well-defined attractors. We are then able to identify the largest value that the Hamiltonian takes over those attractors with the scaled cumulant generating function of our observables, providing a non-linear equivalent to the well-known Donsker-Varadhan formula for jump processes.

On that basis, we prove that chemical reaction networks that are deterministically multistable generically undergo first-order dynamical phase transitions in the vicinity of zero bias. We illustrate that fact through a simple bistable model called the Schl\"ogl model, as well as multistable and unstable generalisations of it, and we make a few surprising observations regarding the stability of deterministic fixed points, and the breaking of ergodicity in the large-volume limit.
\end{abstract}

\maketitle

\section{Introduction}
\label{I}

The fundamental question of statistical physics is whether it is possible to describe a microscopic model, with many components and relatively simple rules, using only a small set of mesoscopic variables, and to determine the emerging laws of the system at that scale. A good but rather trivial example of this can be found in the behaviour of independent random walkers (particles) on a lattice: although the time-evolution of the complete probability distribution of configurations is relatively complicated, one can write a much simpler equation for the distribution of the local average density of particles, under an appropriate rescaling of time and space. That equation takes the form of a Langevin equation, with a drift and a diffusion that are both linear in the density, and a noise whose variance is linear as well. The resulting process is called a linear stochastic diffusion, and is said to be \textit{hydrodynamic} in the sense that the behaviour of the local density is autonomous (i.e. does not depend on higher moments of the density), by analogy with fluid mechanics.

A non-linear generalisation of this is found in the macroscopic fluctuation theory (MFT, \cite{Bertini2006}), which deals with interacting particles close to equilibrium: in essence, if mesoscopic portions of the system get asymptotically close to equilibrium as the size of the system is increased, then the local density of particles obeys a Langevin equation similar to the one mentioned above, but with non-linear drift, diffusion, and noise variance.

The appropriate language to tackle such questions is that of large deviations \cite{DenHollander2008,Touchette20091}, for the same reason that an equilibrium system is characterised by its free energy. Indeed, the process equivalence that we invoke when stating such results is a logarithmic one \cite{Chetrite2014}: the logarithm of the probability distribution of the mesoscopic density profiles of our system, rescaled by the appropriate parameter (volume, or total number of particles), converges in the limit of that parameter going to infinity to the rescaled logarithm of the probability distribution of the corresponding Langevin equation. The argument of the exponential distribution is then generally called the \textit{action}, and for a Markovian process it takes the form of a time integral of a functional of states (density, position) and fluxes (velocity, currents) called the Lagrangian. In the case of diffusions, or of single particles subject to a Gaussian noise with small variance, that Lagrangian is a quadratic function of the fluxes, and has been the topic of many studies in recent years \cite{Chetrite2015,Bouchet2016a,Derrida2018,Tizon-Escamilla2019}. Moreover, in that context, distributions of observables such as time-averaged densities or fluxes are of special interest, as they give information on the way that systems can sustain unlikely fluctuations of those observables, potentially leading to strategies to enhance said fluctuations (e.g. to boost the efficiency of a machine \cite{Verley2014,Gingrich2014}) or suppress (e.g. to prevent catastrophic events \cite{Ragone2018}).

~~

This topic is well-established and well-studied for systems which are rescaled to a continuous-space process, but they make no less sense in situations with an intrinsically discrete geometry. A simple Markov jump process on a finite state-space, for instance, can be interpreted as the evolution equation for the typical density distribution of a large number of independent random walkers, and the probability distribution of that density in the limit of many walkers can be described through the large deviation formalism \cite{0295-5075-82-3-30003}. The resulting process is linear in the density as for diffusions, due to the independence of its components.

In this paper, we consider a specific class of interacting jump processes, which provide a natural nonlinear generalisation of Markov jump processes, namely \textit{chemical reaction networks}. Their deterministic versions are the well-known chemical rate equations systems \cite{Feinberg1987}, which are structurally equivalent to a jump process where the probabilities are replaced by monomials of concentrations. Their behaviour can however be much more complex than their linear counterpart: whereas an irreducible Markov jump process on a finite state-space has a single steady state in the stationary limit, a chemical system can exhibit multistability \cite{Schlogl1972}, limit cycles \cite{Andrieux2008}, or even strange attractors \cite{Olsen1983}. The large deviations of these models, as well as other types of population dynamics, have been a growing subject of interest in the past decade from various communities \cite{Dickman2003,Smith2011,Mielke2016,Rao2016,Assaf2017,Kraaij2018,Agazzi2018} with roots dating back to the 70's \cite{Dickman2003}, although to our knowledge a full theory of their dynamical fluctuations and rare event probabilities is still missing. 

~

In the present work, we describe the large deviations behaviour of generic chemical reaction networks, starting from the chemical master equation \cite{Gillespie1992} which describes their microscopic dynamics.

In section \ref{II}, we introduce those models, along with a few important concepts to analyse them, and we define the mesoscopic variables that will be relevant in the large-volume limit.

In section \ref{III}, we take said large-volume limit, and show that the dynamics of those systems can be described through a path integral with a Lagrangian that can be computed explicitly. That path integral is dominated by a trajectory solving an Euler-Lagrange equation, or equivalently Hamilton's equations for an appropriate Hamiltonian.  This will be done in several different ways, depending on whether one takes the time-derivative of the density or all the separate chemical currents as a flux variable in the Lagrangian, and on whether one puts a bias on some observables through a Lagrange multiplier. A connection will also be made between our approach and the Doi-Peliti formalism \cite{Doi1976,Peliti1985}. This section is meant as a pedagogical overview of the Lagrangian/Hamiltonian formalism for large-volume stochastic processes, and is detailed accordingly. As such, it contains many results that are likely not original, although we have been unable to find references to some of them in the literature (such as for the detailed formalism of section \ref{IIIB}). The focus is however put on the biased version of section \ref{IIIC}, which we do believe to be new in its most general form. The learned reader can safely skip this section or simply skim it for notations. The proofs relating to these results are provided in an Appendix \ref{Appendix}.

Section \ref{IV} contains the main original results of this work: we apply said formalism to compute the scaled cumulant generating (SCGF) function of currents and densities of our models, which is the Legendre transform of the large deviations function of stationary currents and densities. We show that it can be expressed in terms of the maximal value of the biased Hamiltonian defined in section \ref{II} among those it takes on its critical manifolds. We then show that systems whose deterministic equations have several attractors (i.e. that are multistable) generically undergo first-order dynamical phase transitions around small values of the dynamical biases.

In section \ref{V}, we illustrate this last result by exhibiting a few of those dynamical phase transitions in variants of a simple bistable system called the Schl\"ogl model \cite{Schlogl1972}. We remark that, surprisingly, even unstable fixed points of the deterministic dynamics can become stable under bias, and that the ergodicity of the microscopic process, broken by the large-volume limit, is usually restored. We also show that, in some cases where the deterministic chemical equations have stable or unstable regimes depending on their initial condition, a bias can turn the whole system unstable.

Finally, we conclude with a few interesting open questions on the topic.

\section{Definition of the microscopic process and observables}
\label{II}

\subsection{Definition of the process}
\label{IIA}

We will be looking at the standard chemical master equation with mass-action kinetics \cite{Feinberg1987}. Our system contains chemical constituents (or \textit{particles}) of species $X_i$, $i\in[\![0,N[\![$, and a microscopic state $n$ of the system is given by the number of (indistinguishable) particles of each species, which we will write as $n_i$ (we will treat $n$ as a vector with components $n_i$). We assume the system, of volume $V$, is well mixed and has no spatial dimension. The system evolves through \textit{chemical reactions} where a certain set of particles (reactants) is destroyed, and replaced by another set (products), with a certain rate at Poisson-distributed times. This defines a Markov jump process on the state-space $\mathbb{N}^N$ (possibly on a subset of it if all states are not connected dynamically due to conservation laws ; see section \ref{IVC}).

In order to define the process in a convenient way, we will not differentiate reactants and products (as reactions will often be reversible), and we will call all sets of particles occurring on either side of our reactions as \textit{complexes}. A complex $\gamma$ is defined by a set of non-negative integers $\nu^\gamma_i$ called \textit{stoichiometric coefficients}, giving the number of particles of species $X_i$ entering in complex $\gamma$, and the standard notation for the reaction destroying complex $\gamma$ and creating complex $\gamma'$ is then
\begin{equation}
\sum\limits_i \nu^\gamma_i X_i ~~\rightarrow~~ \sum\limits_i \nu^{\gamma'}_i X_i.
\end{equation}
The rate at which this reaction is occurring from state $n$ will be written as $W_{\gamma'\gamma}(n)$, taking the system from $\{n_i\}$ to $\{n_i-\nu^\gamma_i+\nu^{\gamma'}_i\}$. It will be convenient to define the difference between those states to be a generalised divergence
\begin{equation}
\nabla_{\gamma'\gamma}=\nu^{\gamma}-\nu^{\gamma'}
\end{equation}
which acts on the space of reactions (edges of the graph of connected complexes) and yields a variation of particle number. Note that this operator $\nabla$ can be factorised as $\nabla=\nu \mathfrak{d}$, where $\mathfrak{d}$ is the standard discrete divergence on the graph of complexes, acting on reactions and yielding an antisymmetric function on the vertices of that graph (i.e. on complexes), and where $\nu$ is the matrix containing all the stoichiometric coefficients, acting on complexes and yielding particle numbers. We will only need this factorisation in section \ref{IIC} when considering conservation laws.

Moreover, the conjugate of this divergence defines a natural gradient from functions of the species $x$ to antisymmetric functions of the reactions:
\begin{equation}
(\nabla \rho)_{\gamma' \gamma}=\sum\limits_x (\nu^{\gamma'}_x-\nu^{\gamma}_x)\rho_x.
\end{equation}
We use the same notation $\nabla$ as for the divergence, by analogy with the usual continuous case. However, this should not create any ambiguity, as the left and right spaces are different.
We can then perform integrations by parts on functions of reactions: for two functions $\rho_x$ and $\lambda_{\gamma' \gamma}$, we have
\begin{equation}\label{integParts}
\rho \nabla\cdot\lambda=-\lambda\cdot\nabla\rho.
\end{equation}

~~

We can now write the chemical master equation with rates $W_{\gamma'\gamma}(n)$, for a probability distribution $P(n)$:
\begin{equation}
\mathrm{d}_t P(n)=\sum\limits_{\gamma', \gamma}W_{\gamma'\gamma}(n+\nabla_{\gamma', \gamma})P(n+\nabla_{\gamma', \gamma})-W_{\gamma'\gamma}(n)P(n)
\end{equation}
which we can also write algebraically in the bra-ket notation as $\mathrm{d}_t|P\rangle=W|P\rangle$, with the Markov matrix $W$:
\begin{equation}
W=\sum\limits_{\gamma', \gamma,n}W_{\gamma'\gamma}(n)~| n-\nabla_{\gamma', \gamma}\rangle\langle n|-W_{\gamma'\gamma}(n)~| n\rangle\langle n|.
\end{equation}

A common and natural choice for $W_{\gamma'\gamma}(n)$ is the so-called \textit{mass action} prescription, where the dependence in $n$ simply comes from the combinatorial number of ways to choose the reactants, i.e. the product of ${n_i}\choose{\nu^\gamma_i}$ over $\gamma$. since $\nu^\gamma_i!$ is fixed, we may absorb it into the constant prefactor. Moreover, we will later need the prefactor to scale with the volume $V$ with a specific exponent for $V$ large, so for the sake of convenience we will write that term explicitly already. We get
\begin{equation}
W_{\gamma'\gamma}(n)=k_{\gamma'\gamma}\prod_{x}\frac{[n_x]!}{[n_x-\nu_x^\gamma]!}V^{1-\sum_x \nu_x^\gamma},
\end{equation}
where $k_{\gamma'\gamma}$ is called the \textit{kinetic constant} of the reaction, and is independent of $n$ and of $V$.

Our aim will be to describe the fluctuations of this model around its typical behaviour in the $V\rightarrow\infty$ limit. The order of $V$ in each rate has been chosen so that the rates $W(n)$ be all linear in $V$ in the large volume limit, which corresponds to a so-called \textit{hyperbolic} scaling (also known as ballistic scaling). We will see in section\ref{IIIB2} that this scaling is compatible with the standard deterministic mass-action chemical equation systems \cite{Feinberg1987}.

\subsection{Time-additive dynamical observables}
\label{IIB}

When taking the large volume limit, we will reduce the number of dynamical observables in our system: we will not be interested in the complete probability distribution of $n$ and in the complete set of microscopic reactions dependent on the starting state $n$ at every instant, but only in the typical concentration of particles and total chemical currents integrated over an appropriate time window.

Let us therefore consider a single realisation $n(\tau)$ of the microscopic process described above, and let us take a mesoscopic time step $\delta t$, which for now is simply a positive real constant. We can then define the empirical \textit{concentration} $\rho_x$ of species $X$ as
\begin{equation}
\rho_x(t)~V~\delta t=\int_t^{t+\delta t} n_x(\tau)\mathrm{d}\tau
\end{equation}
and the empirical chemical currents $\lambda_{\gamma'\gamma}$ as
\begin{equation}
\lambda_{\gamma'\gamma}(t)~V~\delta t=w_{\gamma'\gamma}\equiv\#\left[\gamma \rightarrow \gamma'\right]_{t,t+\delta t}
\end{equation}
where the last term simply means the number of times reaction $\gamma \rightarrow \gamma'$ happens between $t$ and $t+\delta t$, regardless of the state. Those observable are called \textit{time-additive} because they obey a Chasles relation in time on any history of the system. As a consequence, they are typically linear with respect to the observation time (hence the factors $\delta t$ assumed in the left-hand sides).

Given those definitions, one can check that those quantities are related in the following way:
\begin{equation}
{\mathrm d}_t \rho_x=\sum\limits_{\gamma,\gamma'}(\nu_x^{\gamma'}-\nu_x^{\gamma})\lambda_{\gamma'\gamma}~~~~\mathrm{which~we~write~as}~~~~{\mathrm d}_t \rho=-\nabla\cdot\lambda.
\end{equation}
This is the continuity equation of our chemical system, and it is verified individually by each microscopic realisation.

\subsection{A caveat on conserved quantities}
\label{IIC}

An interesting feature of chemical reaction networks is that they may have many nontrivial conserved quantities: indeed, looking at the continuity equation, we see it involves the operator $\nabla$ which might not be invertible. For any vector $\mathfrak{c}_x$ in the left kernel of $\nabla$, the quantity $\mathfrak{c}\rho$ is conserved:
\begin{equation}
{\mathrm d}_t \mathfrak{c}\rho=-\mathfrak{c}\nabla\cdot\lambda=0.
\end{equation}
We call this equation a \textit{conservation law}, and we will use the same term to designate the vector $\mathfrak{c}$ itself.

The effect of this is that the microscopic state-space of the system is split into many ergodic components, each corresponding to one set of values of all conserved quantities. This also means that the variable $\dot\rho$ is potentially of lower dimension than the number of species, which means that one has to be careful when taking Legendre transforms with respect to it, as we will very often do. In all the following, we will dismiss this issue by implicitly defining $\rho$ on a \textit{reduced state space orthogonal to the conserved quantities}, i.e. one single ergodic component. In such a space, the matrix $\nabla$ is invertible on the concentration side (i.e. to the right for the gradient and to the left for the divergence). We will however have to address the issue in section \ref{IVC} when considering initial conditions that span several ergodic components.

~

We will be interested in the distribution of additive observables in the limit of large volume and small time step, but let us first consider a different case.

\subsection{Aside: long-time large deviations}
\label{IID}

Let us consider for a moment a finite volume $V$ and a single time step $\delta t=t$ which is large with respect to the relaxation time of the system. It is not guaranteed in general that said relaxation time be well defined, due to the possibility of the state-space being infinite, but we will assume that it is (which we can guarantee if the system has a certain type of conservation law, but which is generally true if $P(n)$ is mostly contained in a compact at all times \cite{Feinberg1987}). Let us also assume that our initial distribution has support in only one connected (ergodic) component of the state space.

In this context, let us examine the probability distributions $\mathrm{P}_{t}(\lambda,\rho)$ of chemical currents and concentrations averaged over a single long time step. Given that this is a time-averaged additive observable in a Markov jump process on a finite state space (or a well-controlled infinite one), we know the following \cite{Touchette20091}: the aforementioned distribution follows a large deviation principle with a scale $t$, i.e. we can define a so-called \textit{long-time large deviation function} $g(\lambda,\rho)$ such that
\begin{equation}
\lim\limits_{t  \rightarrow \infty}\left[-\frac{1}{t}\ln\left(\mathrm{P}_{t}(\lambda,\rho)\right)\right]=g(\lambda,\rho),
\end{equation}
which we write as $\mathrm{P}_{t}(\lambda,\rho)\asymp \mathrm{e}^{-tg(\lambda,\rho)}$. The function $g$ is, by construction, positive convex and vanishes only at the stationary value of the currents $\lambda$ and concentrations $\rho$. Moreover, we can define the Legendre transform $E(\mathfrak{s},h)$ of $g(\lambda,\rho)$, with a variable $\mathfrak{s}_{\gamma'\gamma}$ conjugate to $\lambda_{\gamma'\gamma}$ and $h_x$ conjugate to $\rho_x$:
\begin{equation}
E(\mathfrak{s},h)=\mathfrak{s}\cdot\lambda+h\rho-g(\lambda,\rho)~~~~\mathrm{with}~~~~\mathfrak{s}_{\gamma'\gamma}=\partial_{\lambda_{\gamma'\gamma}}g(\lambda,\rho)~~~~\mathrm{and}~~~~h_x=\partial_{\rho_x}g(\lambda,\rho),
\end{equation}
which is also the scaled cumulant generating function (SCGF) of $\lambda$ and $\rho$:
\begin{equation}\label{SCGFt}
\langle \mathrm{e}^{t( \mathfrak{s}\cdot\lambda+h\rho)}\rangle=\mathrm{e}^{t E(\mathfrak{s},h)}
\end{equation}
where $\langle\cdot\rangle$ represents an average over realisations, starting from any initial condition (which does not matter in the long-time limit).
It is a classical result from the Donsker-Varadhan theory of large deviations \cite{Donsker2010a,Donsker1975a,Donsker1976a,Donsker1983a} that the function $E(\mathfrak{s},h)$ is also the \textit{unique largest eigenvalue of a deformed Markov matrix} $W_{\mathfrak{s},h}$, where the non-diagonal entries carry an extra exponential weight $\mathrm{e}^{\mathfrak{s}_{\gamma', \gamma}}$ and the diagonal entries have an extra linear term $\sum_x h_x \rho_x$:
\begin{equation}\label{Wsh}
W_{\mathfrak{s},h}=\sum\limits_{\gamma', \gamma,n}\mathrm{e}^{V\mathfrak{s}_{\gamma', \gamma}}W_{\gamma'\gamma}(n)~| n-\nabla_{\gamma', \gamma}\rangle\langle n|-W_{\gamma'\gamma}(n)~| n\rangle\langle n|+\sum\limits_{x,n} h_x n_x~| n\rangle\langle n|.
\end{equation}

As demonstrated in the Appendix \ref{Appendix}, this biased Markov matrix $W_{\mathfrak{s},h}$ is useful even if the observation time $t$ is small: the role of its largest eigenvalue is lost, but it can still be used as a generator of the biased dynamics which produces the cumulant generating function $E(\mathfrak{s},h)$ in the long-time limit (which in that case will be the same as in \eqref{SCGFt} up to a factor $V$).

Finally, note that in the case where the initial distribution spans several ergodic components $\alpha$ of the state-space, the term with the largest eigenvalues $E_\alpha(\mathfrak{s},h)$ and a non-zero initial probability will exponentially dominate the others. This can lead to first-order phase transitions with respect to $\mathfrak{s}$ or $h$, as those eigenvalues are allowed to cross.

\section{Dynamical large deviations formalism}
\label{III}

In this section, we take the limits $V\rightarrow\infty$ and $\delta t\rightarrow 0$ on our microscopic process, and describe the exponential rate functions (Lagrangians) of the probability distributions of the observables defined in section \ref{IIB}, as well as their Legendre transforms (Hamiltonians), and obtain equations for the corresponding minimisation problems. The proof of the large-volume limit is provided in the Appendix \ref{Appendix}, where we see that it is also essential that, in the limit, $V\delta t\rightarrow\infty$. All of the following derivations also apply to population models with more complex rates (i.e. if $k_{\gamma'\gamma}$ depends on $n$), except for those of section \ref{IVC} which rely on mass-action kinetics.

We will look at three versions of the formalism: one where the flux-type variable of the Lagrangian is the time-derivative of the concentration (standard), one where we keep track of every individual chemical current (detailed), and one where we constrain the dynamics on a certain time-averaged value of the concentrations and currents (biased). In each subsection, we will derive the appropriate large volume large deviation rate function (Lagrangian) and the corresponding cumulant generating function for fluxes (Hamiltonian), and derive the equations of motion that solve the corresponding extremisation problems (i.e. that describe the typical behaviour of the system) along with the appropriate boundary conditions.

\subsection{Standard formalism}
\label{IIIA}

We first describe the standard large deviations formalism for large volume Markov processes, sometimes called the WKB formalism \cite{Dickman2003,Assaf2017} although in this case the low-noise property is proven rather than assumed.

Let us consider the integrated transition rate of our microscopic process, for a finite runtime $t$, between an initial state $n_i$ and a final state $n_f$:
\begin{equation}
\mathrm{P}_t[n_f|n_i]=\langle n_f|\mathrm{e}^{tW}|n_i\rangle.
\end{equation}

If we now observe the process over $K$ mesoscopic time steps $\delta t$, corresponding to a total time $t=K\delta t$, the probability that it is in state $n_k$ at time $k\delta t$ is given by
\begin{equation}
\mathrm{P}_t[\{n_k\}]=\prod\limits_{k=1}^{K}\mathrm{P}_{\delta t}[n_k|n_{k-1}].
\end{equation}
Note that here, the index $k$ corresponds to the time step rather than a chemical species, and every $n_k$ is a full composition vector. The total transition rate between $n_0$ and $n_K$ can then be decomposed over all possible paths:
\begin{equation}\label{microPath}
\mathrm{P}_t[n_K|n_0]=\sum\limits_{k=1}^{K-1}\sum\limits_{n_k=0}^\infty\mathrm{P}_t[\{n_k\}]=\sum\limits_{k=1}^{K-1}\sum\limits_{n_k=0}^\infty\prod\limits_{k=1}^{K}\mathrm{P}_{\delta t}[n_k|n_{k-1}].
\end{equation}

\subsubsection{Standard Lagrangian}
\label{IIIA1}

For an appropriate scaling of time and volume such that $V\delta t\rightarrow\infty$, we expect each mesoscopic transition rate $\mathrm{P}_{\delta t}[n_k|n_{k-1}]$ to have  a large deviation form with a rate which we will call the \textit{Lagrangian} of the process. More precisely, let us define
\begin{equation}\label{Lagr}
\mathcal{L}(\dot\rho,\rho)=-\lim\limits_{V\delta t  \rightarrow \infty}\left[\frac{1}{V\delta t}\ln\Bigl(\mathrm{P}_{\delta t}[V\rho+V\delta t\dot\rho,V\rho]\Bigr)\right].
\end{equation}
Equation \eqref{microPath} can then be rewritten, in the limit $V\rightarrow\infty$, $\delta t\rightarrow 0$, $V\delta t\rightarrow\infty$, $K\rightarrow \infty$, $K\delta t=$cst, as a path integral:
\begin{equation}
\mathrm{P}_t[\rho_t|\rho_0]\asymp\int\mathrm{exp}\left[-V \int_{\tau=0}^{t}\mathcal{L}(\dot\rho(\tau),\rho(\tau))\mathrm{d}\tau\right]\mathcal{D}[\rho].
\end{equation}
Note that this scaling is significantly different from that used in section \ref{IID}, in that the concentrations $\rho$, and in particular the initial and final states, are also scaled in the limit. This means that, even though the number of transitions occurring in a step $\delta t$ is of order $V\delta t\rightarrow\infty$, which is enough to guarantee the exponential scaling of the mesoscopic rates (cf the appendix), the relaxation time of the system becomes of order $V$, and the system does not have time to reach its stationary state.

The last ingredient missing from the picture is the boundary conditions. The previous expression is appropriate when starting and ending at fixes concentrations $\rho_0$ and $\rho_t$, but we may want instead to start from a certain initial distribution, and trace over the final state with a certain cost function. Those boundary conditions have a crucial impact on the long-time behaviour of our systems, as we will see in section \ref{IVB}. Le us therefore consider an initial distribution $P_0(\rho)$ and a final cost $\mathcal{O}_t(\rho)$ of the form
\begin{equation}
P_0(\rho)\asymp\mathrm{e}^{-VU(\rho)}~~~~\mathrm{and}~~~~\mathcal{O}_t(\rho)\asymp\mathrm{e}^{-V\theta(\rho)}
\end{equation}
where $U$ and $\theta$ are non-negative functions that vanish at one point at least (this can be guaranteed by normalising them appropriately). Given that we are interested only in the exponential scaling of probabilities, we do not need to worry about the prefactors, but note that these expressions do include cases where $P_0$ and $\mathcal{O}_t$ do not scale exponentially with $V$: if the scaling is subexponential, then $U=0$ or $\theta=0$, and if the scaling is superexponential they should be replaced by theta functions centred at their maxima.

Once those terms have been taken into account, the path integral representation of the process becomes
\begin{equation}\label{Path}
\langle\mathcal{O}_t\rangle_{P_0}\asymp\int\mathrm{exp}\left[-V\left( \int_{\tau=0}^{t}\mathcal{L}(\dot\rho(\tau),\rho(\tau))\mathrm{d}\tau+U(\rho_0)+\theta(\rho_t) \right)\right]\mathcal{D}[\rho].
\end{equation}

\subsubsection{Equations of motion}
\label{IIIA2}

Considering that equations \eqref{Path} is a sum over exponentials with a large exponent, the terms dominating the path integral corresponds to the trajectories which minimise the rate of the exponential (or \textit{action}). By taking a functional derivative of the action, we find the deterministic equations of motion describing the typical behaviour of our system.

~

The differentiation of the action with respect to $\rho_\tau$ at every time $\tau$ gives:
\begin{equation}
\int_{\tau=0}^{t}\left(\partial_{\rho}\mathcal{L}~\delta\rho_\tau+\partial_{\dot\rho}\mathcal{L}~\delta\dot\rho_\tau\right)\mathrm{d}\tau+\partial_{\rho}U(\rho_0)\delta\rho_0+\partial_{\rho}\theta(\rho_t)\delta\rho_t=0
\end{equation}
An integration by parts of the second term yields
\begin{equation}
\int_{\tau=0}^{t}\left(\partial_{\rho}\mathcal{L}-\frac{\mathrm{d}}{\mathrm{d}t}\partial_{\dot\rho}\mathcal{L}\right)\delta\rho_\tau~\mathrm{d}\tau+\left[ \partial_{\dot\rho}\mathcal{L}~\delta\rho_\tau\right]_0^t+\partial_{\rho}U(\rho_0)\delta\rho_0+\partial_{\rho}\theta(\rho_t)\delta\rho_t=0.
\end{equation}
Canceling the integrated term will yield the standard Euler-Lagrange equation
\begin{equation}\label{Euler}\boxed{
\partial_\rho\mathcal{L}-\frac{\mathrm{d}}{\mathrm{d}t}\partial_{\dot\rho}\mathcal{L}=0.
}
\end{equation}
and the boundary terms fix the boundary conditions:
\begin{equation}\label{Bound}
\partial_{\dot\rho}\mathcal{L}(\rho_0,\dot\rho_0)=\partial_{\rho}U(\rho_0)~~~~\mathrm{and}~~~~\partial_{\dot\rho}\mathcal{L}(\rho_t,\dot\rho_t)=-\partial_{\rho}\theta(\rho_t).
\end{equation}

\subsubsection{Standard Hamiltonian and Hamilton's equations}
\label{IIIA3}

Let us start once more from eq.\eqref{microPath}, and let us define a rescaled cumulant generating function (i.e. a log-Laplace transform) of $\mathrm{P}_{\delta t}[n|n_i]$ with respect to $n-n_i$, with a parameter $f_x$ conjugate to $n_x$, which we call the standard \textit{Hamiltonian}:
\begin{equation}
\mathcal{H}(f,n_i)=\frac{1}{V\delta t}\ln\Bigl(\sum_n\mathrm{e}^{(n-n_i)f}\langle n|\mathrm{e}^{\delta tW}|n_i\rangle\Bigr).
\end{equation}
Since the difference in state only depends on the integrated chemical currents through the continuity equation, this generating function can be rewritten as
\begin{equation}
\mathcal{H}(f,n_i)=\frac{1}{V\delta t}\ln\Bigl(\sum_n\langle n|\mathrm{e}^{\delta tW_{\nabla f,0}}|n_i\rangle\Bigr),
\end{equation}
with the biased generator $W_{\sigma,h}$ defined in section \ref{IID}.

In the large volume and small time step limit, with $n_i=V\rho$, this expression simply becomes
\begin{equation}\label{Ham}\boxed{
\mathcal{H}(f,\rho)=\sum\limits_{\gamma,\gamma'} k_{\gamma'\gamma}\rho^{\nu^\gamma}\left(\mathrm{e}^{(\nu^{\gamma'}-\nu^\gamma)f}-1\right)
}
\end{equation}
where $\rho^{\nu^\gamma}=\prod\rho_x^{\nu^\gamma_x}$, as detailed in the appendix \ref{Appendix}. Moreover, the Laplace transform becomes a Legendre transform, which means that the Hamiltonian is related to the Lagrangian through
\begin{equation}
\mathcal{H}(f,\rho)=f\dot\rho-\mathcal{L}(\dot\rho,\rho)~~~~~\mathrm{with}~~~~~f=\partial_{\dot\rho}\mathcal{L}
\end{equation}
hence its name. Note that $\mathcal{H}$ is explicit even though $\mathcal{L}$ is not in general.

~

The Euler-Lagrange equations can be recast in terms of the Hamiltonian. Let us consider eq.\eqref{Euler}, and define $f=\partial_{\dot\rho}\mathcal{L}$. The Euler-Lagrange equation becomes
\begin{equation}
\partial_\rho\mathcal{L}=\frac{\mathrm{d}}{\mathrm{d}t}\partial_{\dot\rho}\mathcal{L}=\dot f.
\end{equation}
Defining now $\mathcal{H}=f\dot\rho-\mathcal{L}$, so that $\partial_\rho\mathcal{L}=-\partial_\rho\mathcal{H}$ and $\dot\rho=\partial_{f}\mathcal{H}$, the Euler-Lagrange equation simply becomes one of the standard Hamilton equations:
\begin{equation}\label{Hamilton}\boxed{
\dot f=-\partial_\rho\mathcal{H}~~~~\mathrm{with}~~~~\dot\rho=\partial_{f}\mathcal{H}.
}
\end{equation}
The boundary conditions simply translate to
\begin{equation}\label{HamBound}
f_0=\partial_{\rho}U(\rho_0)~~~~\mathrm{and}~~~~f_t=-\partial_{\rho}\theta(\rho_t)
\end{equation}
which are implicit equations that define a manifold in phase space at each time. Let us also remark that the value of $\mathcal{H}$ is conserved by that dynamics:
\begin{equation}
\dot{\mathcal{H}}=\dot{\rho}~\partial_\rho\mathcal{H}+\dot{f}~\partial_f\mathcal{H}=0.
\end{equation}

\subsubsection{Two-fields picture}
\label{IIIA4}

In the specific case of mass-action kinetics, we can transform the standard Hamiltonian $\mathcal{H}$ to a form closer to the famous Doi-Peliti action \cite{Doi1976,Peliti1985,Dickman2003}, sometimes called Liouville functional \cite{Smith2018a}, which will be convenient for certain computations below.

Let us define two new variables (fields) $\psi_x=\mathrm{e}^{-f_x}\rho_x$ and $\phi_x=\mathrm{e}^{f_x}$, such that $\rho_x=\phi_x\psi_x$ and $f_x=\ln(\phi_x)$. In term of those variables, the Hamiltonian \eqref{Ham} becomes
\begin{equation}\label{twoHam}\boxed{
\mathscr{H}\left(\phi_x,\psi_x\right)=\sum\limits_{\gamma,\gamma'} k_{\gamma'\gamma}\psi^{\nu^\gamma}\left(\phi^{\nu^{\gamma'}}-\phi^{\nu^{\gamma}}\right)=\langle \phi^{\nu}|K|\psi^{\nu}\rangle
}
\end{equation}
where $\psi^\nu$ and $\phi^\nu$ are vectors in the space of complexes, and $K$ is the Markov matrix containing the kinetic rates $k$. Moreover, using eq.\eqref{Hamilton}, the dynamics of those two variables is given by
\begin{align}
\dot\phi_x&=\phi_x\dot f_x=-\phi_x\sum\limits_{\gamma,\gamma'} \frac{\nu^\gamma_x}{\rho_x}k_{\gamma'\gamma}\rho^{\nu^\gamma}\left(\mathrm{e}^{(\nu^{\gamma'}-\nu^\gamma)f}-1\right)=-\sum\limits_{\gamma,\gamma'} \frac{\nu^\gamma_x}{\psi_x}k_{\gamma'\gamma}\psi^{\nu^\gamma}\left(\phi^{\nu^{\gamma'}}-\phi^{\nu^{\gamma}}\right)=-\partial_{\psi_x}\mathscr{H}\nonumber\\
\dot\psi_x&=\frac{\dot\rho_x}{\phi_x}-\psi_x\dot f_x=\frac{1}{\phi_x}\sum\limits_{\gamma,\gamma'} k_{\gamma'\gamma}\rho^{\nu^\gamma}\left(\nu^{\gamma'}_x-\nu^\gamma_x\right)\mathrm{e}^{(\nu^{\gamma'}-\nu^\gamma)f}+\psi_x\sum\limits_{\gamma,\gamma'} \frac{\nu^\gamma_x}{\rho_x}k_{\gamma'\gamma}\rho^{\nu^\gamma}\left(\mathrm{e}^{(\nu^{\gamma'}-\nu^\gamma)f}-1\right) \nonumber\\
&=\frac{1}{\phi_x}\sum\limits_{\gamma,\gamma'} k_{\gamma'\gamma}\psi^{\nu^\gamma}\left(\nu^{\gamma'}_x\phi^{\nu^{\gamma'}}-\nu^{\gamma}_x\phi^{\nu^{\gamma}}\right)=\partial_{\phi_x}\mathscr{H}\nonumber \nonumber
\end{align}
which is to say that they obey the Hamilton equations with the appropriate Hamiltonian. The change of variables between $\{f,\rho\}$ and $\{\phi,\psi\}$ is therefore canonical. The corresponding boundary conditions become
\begin{equation}\label{DoiBound}
\phi_0=\exp\left[\psi_0^{-1}\partial_{\phi}U(\phi_0\psi_0)\right]~~~~\mathrm{and}~~~~\phi_t=\exp\left[-\psi_t^{-1}\partial_{\phi}\theta(\phi_t\psi_t)\right].
\end{equation}

The usefulness of this so-called \textit{two-fields} variables $\{\phi,\psi\}$, as opposed to the \textit{density-phase} variables $\{f,\rho\}$, is twofold. First, we may in certain cases use the algebraic structure of eq.\eqref{twoHam} to get information on the stationary states of the system \cite{Anderson2010a}. Second, it turns out to be more appropriate for certain types of boundary conditions, as we will see in section \ref{IVC}. Note that, unlike the standard Doi-Peliti formalism, all variables here are real-valued, although it makes no practical difference when dealing with the corresponding path integrals.

\subsection{Detailed formalism}
\label{IIIB}

In practice, the Lagrangian $\mathcal{L}(\dot\rho,\rho)$ is in general difficult to compute explicitly, due to the potentially large number of trajectories resulting in the same $\dot\rho$. We will see in section \ref{V} that this can still be achieved for systems with a single chemical species \cite{Elgart2004,Assaf2010}, but in general there is a more appropriate approach: rather than only keeping track of the variation of concentrations, we can differentiate the mesoscopic transition rates depending on the rate of each individual reaction. This corresponds to the so-called \textit{2.5 level} of large deviations \cite{Barato2015}, where the highest possible mesoscopic detail is kept from microscopic trajectories. Several examples of the Lagrangians and Hamiltonians thus produced can be found in \cite{Kraaj} in a different context.

Define $\hat{\mathrm{P}}_{\delta t}[w_{\gamma' \gamma}|n_i]$ to be the probability, from $n_i$, to perform $w_{\gamma' \gamma}$ times reaction $\gamma\rightarrow\gamma'$ during a time $\delta t$. This rate is such that
\begin{equation}\label{detMicroPath}
\mathrm{P}_{\delta t}[n_f|n_i]=\sum\limits_{\gamma', \gamma}\sum\limits_{w_{\gamma' \gamma}}\hat{\mathrm{P}}_{\delta t}[w_{\gamma' \gamma}|n_i]~\mathbb{I}(\nabla\cdot w=n_f-n_i)
\end{equation}
where $\mathbb{I}$ is the indicator function (i.e. a Kronecker delta), used for typographic convenience.

\subsubsection{Detailed Lagrangian}
\label{IIIB1}

This new decomposition allows us to define a new large deviation function, which we will call the \textit{detailed Lagrangian} of our process:
\begin{equation}
L(\lambda,\rho)=-\frac{1}{V\delta t}\ln\Bigl(\hat{\mathrm{P}}_{\delta t}[V\delta t\lambda |V\rho]\Bigr).
\end{equation}
Equation \eqref{microPath} can be rewritten as a new path integral, with the indicator function becoming a continuous delta function:
\begin{equation}
\mathrm{P}_t[\rho_t|\rho_0]\asymp\int_{\rho_0}\mathrm{exp}\left[-V \int_{\tau=0}^{t}L(\lambda(\tau),\rho(\tau))\mathrm{d}t\right]~\delta(\dot\rho+\nabla\cdot\lambda)~\mathcal{D}[\lambda]
\end{equation}
and including the boundary conditions yields
\begin{equation}\label{detPath}
\langle\mathcal{O}_t\rangle_{P_0}\asymp\int\mathrm{exp}\left[-V\left( \int_{\tau=0}^{t}L(\lambda(\tau),\rho(\tau))\mathrm{d}\tau+U(\rho_0)+\theta(\rho_t) \right)\right]~\delta(\dot\rho+\nabla\cdot\lambda)~\mathcal{D}[\lambda].
\end{equation}
The advantage of this approach is that, unlike $\mathcal{L}$, the detailed Lagrangian $L$ can always be computed explicitly, as shown in the Appendix \ref{Appendix}:
\begin{equation}\label{detLag}\boxed{\boxed{
L(\lambda,\rho)=\sum\limits_{\gamma,\gamma'}\lambda_{\gamma'\gamma}\ln(\lambda_{\gamma'\gamma}/k_{\gamma'\gamma}\rho^{\nu^\gamma})- \lambda_{\gamma'\gamma}+k_{\gamma'\gamma}\rho^{\nu^\gamma}.
}}
\end{equation}
This is the first important result of this paper. Note the similarity with the long-time large deviations function of currents and densities in a Markov jump process \cite{0295-5075-82-3-30003}, which is due to the fact that transition events have Poissonian distributions in both cases, as is demonstrated in the Appendix \ref{Appendix}. In that respect, the large volume limit of chemical reaction networks is a \textit{natural nonlinear version of Markov jump processes}, in the same way that interacting diffusions, as described by the macroscopic fluctuation theory \cite{Bertini2006}, are a natural nonlinear version of Fokker-Planck equations. It was brought to our attention that a similar result can be found in \cite{Monthus2018} in a case where the transitions are limited to the exchange of a single particle.

The standard Lagrangian can then be obtained through the contraction formula
\begin{equation}\label{contrLagr}
\mathcal{L}(\dot\rho,\rho)=\min_{\dot\rho=-\nabla\cdot\lambda} L(\lambda,\rho).
\end{equation}

\subsubsection{Equations of motion}
\label{IIIB2}

Deriving the equations of motion for the detailed Lagrangian $L$ requires more care, as we have to take the conservation relation into account. This is done by introducing a Lagrange multiplier $\xi_x(t)$ conjugate to the conserved quantity $(\dot\rho+\nabla\cdot\lambda)_x(t)$, so that $\rho$ and $\lambda$ become independent variables. The bulk minimisation of the path integral yields three independent parts
\begin{equation}
\delta \left[L+\xi(\dot\rho+\nabla\cdot\lambda)\right]=\left[\partial_\rho L~\delta\rho+\xi~\delta\dot\rho\right]+\left[\partial_{\lambda}L\cdot\delta\lambda+\xi\nabla\cdot\delta\lambda\right]+\left[\dot\rho+\nabla\cdot\lambda\right]\delta \xi=0.
\end{equation}
Each part has to vanish on its own. We can do an integration by parts on time in the first term, and on space in the second using eq.\eqref{integParts}, which gives two equations in addition to the conservation of matter from the third term:
\begin{equation}
\partial_\rho L=\dot \xi~~~~\mathrm{and}~~~~\partial_{\lambda}L=\nabla \xi.
\end{equation}
Combining the two yields the detailed Euler-Lagrange equation
\begin{equation}\label{detEuler}\boxed{
\nabla\partial_\rho L-\frac{\mathrm{d}}{\mathrm{d}t}\partial_{\lambda}L=0.
}
\end{equation}
Moreover, the boundary terms from the partial integration of $f~\delta\dot\rho$, combined with the differentials of $U_0$ and $\theta_t$, yield the boundary conditions
\begin{equation}
\xi_0=\partial_{\rho}U(\rho_0)~~~~\mathrm{and}~~~~\xi_t=-\partial_{\rho}\theta(\rho_t)
\end{equation}
so that
\begin{equation}\label{detBound}
\partial_{\lambda}L(\lambda_0,\rho_0)=\nabla\partial_{\rho}U(\rho_0)~~~~\mathrm{and}~~~~\partial_{\lambda}L(\lambda_t,\rho_t)=-\nabla\partial_{\rho}\theta(\rho_t)
\end{equation}
Note the similarity with the standard Hamiltonian boundary conditions. However, in this case, $\xi$ is not the variable of the Hamiltonian, as we will see in the next section.

Those equations of motion must be verified by the typical trajectories of the system, in particular if we start from a certain concentration $\rho_0$ and simply trace over the final concentration by taking $\theta_t=0$. Since we have an explicit expression for $L$, we could inject it in \eqref{detEuler}, but that is not necessary: we know that $L$ is positive by construction and cancels only at
\begin{equation}
\lambda_{\gamma'\gamma}=k_{\gamma'\gamma}\rho^{\nu^\gamma}.
\end{equation}
This, combined with the conservation equation, yields the standard system of coupled first order differential equations describing the evolution of a dilute well-mixed chemical solution with mass-action kinetics:
\begin{equation}
\dot\rho_x=\sum\limits_{\gamma,\gamma'} (\nu_x^{\gamma'}-\nu_x^\gamma)k_{\gamma'\gamma}\rho^{\nu^\gamma}.
\end{equation}
Moreover, since this solution gives a value for $\dot\rho$ that only depends on $\rho$, this is also the solution of equation \eqref{Euler}.

\subsubsection{Detailed Hamiltonian and Hamilton's detailed equations}
\label{IIIB3}

As with the Lagrangian, we can also define a \textit{detailed Hamiltonian}, starting from eq.\eqref{detMicroPath} and introducing a variable $\sigma_{\gamma'\gamma}$ conjugate to $w_{\gamma'\gamma}$:
\begin{equation}
H(\sigma,n)=\frac{1}{V\delta t}\ln\Bigl(\sum_{\gamma',\gamma}\mathrm{e}^{\sigma_{\gamma' \gamma} w_{\gamma' \gamma}}\hat{\mathrm{P}}_{\delta t}[w_{\gamma' \gamma}|n_i]\Bigr)
\end{equation}
which, as above, can be rewritten as
\begin{equation}
H(\sigma,n)=\frac{1}{V\delta t}\ln\Bigl(\sum_n\langle n|\mathrm{e}^{\delta tW_{\sigma,0}}|n_i\rangle\Bigr)
\end{equation}
which we recognise as a generalisation of the standard Hamiltonian, where the \textit{virtual forces} are not necessarily of gradient form. We use here the letter $\sigma$ instead of $\mathfrak{s}$ for the variable conjugate to $\lambda$, in order to avoid confusion later when we introduce both a bias and a conjugation on $\lambda$ simultaneously. Even though those variables occupy the same place in the definition of the Hamiltonians, they represent different physical objects.

In the large volume and small time step limit, with $n_i=V\rho$ and $w=V\delta t\lambda$, this expression simply becomes
\begin{equation}\label{detHam}\boxed{
H(\sigma,\rho)=\sum\limits_{\gamma,\gamma'} k_{\gamma'\gamma}\rho^{\nu^\gamma}\left(\mathrm{e}^{\sigma_{\gamma'\gamma}}-1\right),
}
\end{equation}
and we can check that $H$ is indeed the Legendre transform of $L$:
\begin{equation}
H(\sigma,\rho)=\sigma\cdot\lambda-L(\lambda,\rho)~~~~~\mathrm{with}~~~~~\sigma_{\gamma'\gamma}=\partial_{\lambda_{\gamma'\gamma}}L.
\end{equation}

The equivalent of the contraction formula \eqref{contrLagr} for Hamiltonians is a simple reduction of variables
\begin{equation}\label{contrHam}
\mathcal{H}(f,\rho)=H(\nabla f,\rho)
\end{equation}
which is one reason to favour Hamiltonians over Lagrangians for explicit computations.

Note that this contraction is compatible with the continuity equation $\dot\rho=-\nabla\cdot\lambda$, as setting $\sigma=\nabla f$ leads to 
\begin{equation}
\sigma\cdot\lambda=-f\nabla\cdot\lambda=f\dot\rho,
\end{equation}
which is to say that the Legendre scalar $\sigma\cdot\lambda$ is conserved through its contraction to $f\dot\rho$.

~~

Starting from eq.\eqref{detEuler}, and defining $\sigma=\partial_{\lambda}L$, we find
\begin{equation}
\nabla\partial_\rho L=\frac{\mathrm{d}}{\mathrm{d}t}\partial_{\lambda}L=\dot \sigma.
\end{equation}
Defining $H=\sigma\cdot\lambda-L$, so that $\partial_\rho L=-\partial_\rho H$ and $\dot\rho=-\nabla\cdot\partial_{\sigma}H$, we get the detailed Hamilton equations:
\begin{equation}\label{detHamilton}\boxed{
\dot \sigma=-\nabla\partial_\rho H~~~~\mathrm{with}~~~~\dot\rho=-\nabla\cdot\partial_{\sigma}H.
}
\end{equation}
The boundary conditions become
\begin{equation}\label{detHamBound}
\sigma_0=\nabla\partial_{\rho}U(\rho_0)~~~~\mathrm{and}~~~~\sigma_t=-\nabla\partial_{\rho}\theta(\rho_t)
\end{equation}
The value of $H$ is also a constant of motion:
\begin{equation}
\dot{H}=\dot{\rho}~\partial_\rho H+\dot{\sigma}~\partial_\sigma H=-(\nabla\cdot\partial_{\sigma}H)\partial_\rho H-(\nabla\partial_\rho H)\partial_\sigma H=0
\end{equation}
through the integration by parts of one of the terms.

All of the equations of this section simply reduce to those for the standard Hamiltonian by setting $\sigma=\nabla f$, which is compatible with the boundary conditions.

\subsection{Biased formalism}
\label{IIIC}

The aim of this work is ultimately to describe the large deviations of our systems when conditioned on certain values of the currents $\lambda$ and densities $\rho$. This can be formally achieved by constraining all the equations of the previous section on the time-average of those observables. However, as we saw in section \ref{IID}, it can be preferable in practice to look at generating functions instead.

\subsubsection{Detailed biased dynamics}
\label{IIIC1}

As shown in the Appendix \ref{Appendix}, the generating function of the cumulants of those observables can be obtained by replacing $W$ by $W_{\mathfrak{s},h}$, as defined in eq.\eqref{Wsh}, in every term of equation \eqref{microPath}. This simply leads to an equation of the exact same form as \eqref{detPath} but with $L$ replaced by a new \textit{biased} Lagrangian
\begin{equation}\label{detBiasLag}\boxed{
L_{\mathfrak{s},h}(\lambda,\rho)=L(\lambda,\rho)-\mathfrak{s}\cdot\lambda-h \rho
}
\end{equation}
i.e.
\begin{equation}\label{detBiasPath}
\langle \mathcal{O}_t\mathrm{e}^{tV( \mathfrak{s}\cdot\lambda+h\rho)}\rangle_{P_0}\!\asymp\!\int\!\exp\!\left[-V\!\left( \int_{\tau=0}^{t}L_{\mathfrak{s},h}(\lambda(\tau),\rho(\tau))\mathrm{d}\tau+U(\rho_0)+\theta(\rho_t) \right)\right]\delta(\dot\rho+\nabla\!\cdot\!\lambda)~\mathcal{D}[\lambda].
\end{equation}

The detailed Euler-Lagrange equation \eqref{detEuler} still applies directly to the biased Lagrangian $L_{\mathfrak{s},h}$, from the same calculations. Those equations can be recast in terms of the detailed Lagrangian $L$ as
\begin{equation}
\partial_\rho L=\dot \xi+h~~~~\mathrm{and}~~~~\partial_{\lambda}L=\nabla \xi+\mathfrak{s}.
\end{equation}
Combining the two, for $\mathfrak{s}$ constant in time, yields the biased Euler-Lagrange equation
\begin{equation}\label{biasEuler}\boxed{
\nabla\partial_\rho L-\frac{\mathrm{d}}{\mathrm{d}t}\partial_{\lambda}L=\nabla h.
}
\end{equation}
Note that the dependence on $\mathfrak{s}$ is not apparent from this equation, but can be seen in the boundary conditions
\begin{equation}
\partial_{\lambda}L_{\mathfrak{s},h}(\lambda_0,\rho_0)=\nabla\partial_{\rho}U(\rho_0)~~~~\mathrm{and}~~~~\partial_{\lambda}L_{\mathfrak{s},h}(\lambda_t,\rho_t)=-\nabla\partial_{\rho}\theta(\rho_t)
\end{equation}
which become, in terms of $L$:
\begin{equation}
\partial_{\lambda}L(\lambda_0,\rho_0)=\mathfrak{s}+\nabla\partial_{\rho}U(\rho_0)~~~~\mathrm{and}~~~~\partial_{\lambda}L(\lambda_t,\rho_t)=\mathfrak{s}-\nabla\partial_{\rho}\theta(\rho_t)
\end{equation}

~~

Defining the biased Hamiltonian is straightforward, as biasing the microscopic dynamics $W$ is an associative operation: $(W_{\mathfrak{s},h})_{\sigma,0}=W_{\mathfrak{s}+\sigma,h}$. Through computations shown in the Appendix \ref{Appendix}, we find
\begin{equation}\label{detBiasHam}\boxed{
H_{\mathfrak{s},h}(\sigma,\rho)=H(\sigma+\mathfrak{s},\rho)+h\rho
}
\end{equation}
which we can check to be the Legendre transform of $L_{\mathfrak{s},h}$:
\begin{equation}
H_{\mathfrak{s},h}(\sigma,\rho)=\sigma\cdot\lambda-L_{\mathfrak{s},h}(\lambda,\rho)~~~~~\mathrm{with}~~~~~\sigma_{\gamma'\gamma}=\partial_{\lambda_{\gamma'\gamma}}L_{\mathfrak{s},h}.
\end{equation}
The Hamilton equations for $H_{\mathfrak{s},h}$ are the same as those for the detailed Hamiltonian, and can be recast as
\begin{equation}\label{biasHamilton}\boxed{
\dot \sigma=-\nabla\partial_\rho H(\sigma+\mathfrak{s},\rho)-\nabla h~~~~\mathrm{with}~~~~\dot\rho=-\nabla\cdot\partial_{\sigma}H(\sigma+\mathfrak{s},\rho)
}
\end{equation}
with the same boundary conditions
\begin{equation}\label{biasHamBound}
\sigma_0=\nabla\partial_{\rho}U(\rho_0)~~~~\mathrm{and}~~~~\sigma_t=-\nabla\partial_{\rho}\theta(\rho_t)
\end{equation}
The value of $H_{\mathfrak{s},h}$ is a constant of motion for the same reason as before. Note that, in the case where $h=0$, the value of $H$ is also constant.

\subsubsection{Contraction of the currents and two-fields picture}
\label{IIIC2}

We saw earlier that the detailed Lagrangian has the advantage of being explicit which the standard Lagrangian usually is not. It has however a serious disadvantage in practice: the set of currents $\lambda$ is of much higher dimension than the set of velocities $\dot\rho$. This also applies to the conjugate quantities $f$ and $\sigma$, although all Hamiltonians are in fact explicit, which makes $\mathcal{H}$ preferable to $H$. The latter is useful in defining the biased Hamiltonian $H_{\mathfrak{s},h}$, but for the same reason it makes sense to define a contracted biased Hamiltonian $\mathcal{H}_{\mathfrak{s},h}(f,\rho)$ with fewer variables.

To that effect, let us formally define a contracted biased Lagrangian $\mathcal{L}_{\mathfrak{s},h}$ by contraction of $L_{\mathfrak{s},h}$:
\begin{equation}
\mathcal{L}_{\mathfrak{s},h}(\dot\rho,\rho)=\min_{\dot\rho=-\nabla\cdot\lambda} L_{\mathfrak{s},h}(\lambda,\rho).
\end{equation}
This cannot be written explicitly in general, but this leads to the following contraction of the Hamiltonians:
\begin{equation}
\mathcal{H}_{\mathfrak{s},h}(f,\rho)=H_{\mathfrak{s},h}(\nabla f,\rho)=H(\mathfrak{s}+\nabla f,\rho)+h\rho
\end{equation}
so that
\begin{equation}\boxed{
\mathcal{H}_{\mathfrak{s},h}(f,\rho)=\sum\limits_{\gamma,\gamma'} k_{\gamma'\gamma}\rho^{\nu^\gamma}\left(\mathrm{e}^{(\nu^{\gamma'}-\nu^\gamma)f+\mathfrak{s}_{\gamma'\gamma}}-1\right)+\sum\limits_x h_x \rho_x.
}
\end{equation}
Notice that this contraction is compatible with the previous Hamilton equations by setting $\sigma=\nabla f$ and removing the $\nabla$ (remember that, as stated in section \ref{IIC}, we are implicitly working in a space where $\nabla$ is invertible on the concentration side), which simply yields
\begin{equation}\label{contrBiasHam}\boxed{
\dot f=-\partial_\rho H(\nabla f+\mathfrak{s},\rho)-h=-\partial_\rho \mathcal{H}_{\mathfrak{s},h}~~~~\mathrm{with}~~~~\dot\rho=\partial_{f}H(\nabla f+\mathfrak{s},\rho)=\partial_{f}\mathcal{H}_{\mathfrak{s},h}
}
\end{equation}
with boundary conditions
\begin{equation}
f_0=\partial_{\rho}U(\rho_0)~~~~\mathrm{and}~~~~f_t=-\partial_{\rho}\theta(\rho_t).
\end{equation}
As always, the value of this Hamiltonian is constant along the solutions of those equations.

~

As in section \ref{IIIA4}, we can define $\psi_x=\mathrm{e}^{-f_x}\rho_x$ and $\phi_x=\mathrm{e}^{f_x}$, such that the Hamiltonian becomes
\begin{equation}\label{twoBiasHam}\boxed{
\mathscr{H}_{\mathfrak{s},h}\left(\phi_x,\psi_x\right)=\sum\limits_{\gamma,\gamma'} k_{\gamma'\gamma}\psi^{\nu^\gamma}\left(\mathrm{e}^{\mathfrak{s}_{\gamma'\gamma}}\phi^{\nu^{\gamma'}}-\phi^{\nu^{\gamma}}\right)+\sum\limits_x h_x \phi_x\psi_x=\langle \phi^{\nu}|K_{\mathfrak{s}}|\psi^{\nu}\rangle+\langle \phi|h|\psi\rangle
}
\end{equation}
where $K_{\mathfrak{s}}$ is the biased Markov matrix containing the kinetic rates $k$. Through the same calculations as before, we can check the Hamilton equations for those variables
\begin{equation}\boxed{
\dot\phi_x=-\partial_{\psi_x}\mathscr{H}_{\mathfrak{s},h}~~~~\mathrm{and}~~~~\dot\psi_x=\partial_{\phi_x}\mathscr{H}_{\mathfrak{s},h}
}
\end{equation}
with boundary conditions
\begin{equation}\label{biasDoiBound}
\phi_0=\exp\left[\psi_0^{-1}\partial_{\phi}U(\phi_0\psi_0)\right]~~~~\mathrm{and}~~~~\phi_t=\exp\left[-\psi_t^{-1}\partial_{\phi}\theta(\phi_t\psi_t)\right].
\end{equation}

\section{Stationary large deviations and dynamical phase transitions}
\label{IV}

Now that the formalism has been set up, we can examine what happens in the long-time limit in order to obtain the SCGF $E(\mathfrak{s},h)$, as defined in section \ref{IID} but with a term $V$ factored out:
\begin{equation}
\langle \mathrm{e}^{tV(\mathfrak{s}\cdot\lambda+h\rho)}\rangle_{P_0}\sim\mathrm{e}^{tVE(\mathfrak{s},h)}~~~~\mathrm{for}~~~~t\rightarrow \infty,
\end{equation}
without having to solve the microscopic eigenvalue problem, which is potentially infinite-dimensional.

As we saw in the previous section, this expression is a special case of the path integral \eqref{detBiasPath} with $\theta_t=0$, which we may write in terms of $\mathcal{L}_{\mathfrak{s},h}$ instead, and is dominated by the term solving the corresponding Euler-Lagrange equations. The value of the dominating term is then given by
\begin{equation}
\langle \mathrm{e}^{tV(\mathfrak{s}\cdot\lambda+h\rho)}\rangle_{P_0}\sim\exp\!\left[-V\!\left( \int_{\tau=0}^{t}\mathcal{L}_{\mathfrak{s},h}(\dot\rho(\tau),\rho(\tau))\mathrm{d}\tau+U(\rho_0)\right)\right]
\end{equation}
along the solution. Rewriting this in terms of the Hamiltonian, we get
\begin{equation}\label{HamSol}
\langle \mathrm{e}^{tV(\mathfrak{s}\cdot\lambda+h\rho)}\rangle_{P_0}\sim\exp\!\left[V\!\left( \int_{\tau=0}^{t}\left(\mathcal{H}_{\mathfrak{s},h}(f(\tau),\rho(\tau))-f\dot\rho\right)\mathrm{d}\tau+U(\rho_0)\right)\right]
\end{equation}
along the solution of the Hamilton equations $\{\rho^\star(\tau),f^\star(\tau)\}$. If the solution is not unique, the one which gives the largest value of the argument of the exponential dominates.

We saw that the value of $\mathcal{H}_{\mathfrak{s},h}$ is conserved along this solution, so that it comes out of the time integral. Moreover, the term $U(\rho_0)$ is not extensive in time, so that it is negligible in the long time limit, though it does appear implicitly through $f^\star$ and $\rho^\star$, which depend on it. We are left with the \textit{kinetic term} $\int f\dot\rho~\mathrm{d}\tau$ which may or may not be extensive in time. We will assume that it is not, which we will support in a specific case in section \ref{IVB}. Under that assumption, we finally get
\begin{equation}
\langle \mathrm{e}^{tV(\mathfrak{s}\cdot\lambda+h\rho)}\rangle_{P_0}\sim\exp\!\left[tV\mathcal{H}_{\mathfrak{s},h}(f^\star,\rho^\star)\right].
\end{equation}

This means that, assuming that the kinetic term is not time-extensive, the SCGF $E(\mathfrak{s},h)$ is equal to the largest value of the biased Hamiltonian $\mathcal{H}_{\mathfrak{s},h}$ along solutions of the Hamilton equations.

~~

In the rest of this section, we will describe general features of those solutions including their attractors, and the consequences of choosing different initial conditions. We will then prove that \textit{multistable} systems (i.e. that have more than one attractor) generically undergo a first-order dynamical phase transition when the bias crosses $0$.

\subsection{Boundary conditions}
\label{IVA}

The choice of boundary conditions can be a determining factor in how the system under consideration fluctuates. Since we are mostly interested in the SCGF defined above, we will restrict ourselves to a flat final condition, i.e.
\begin{equation}
\theta(\rho_t)=0~~~~\mathrm{so~that}~~~~f_t=0~~~~\mathrm{or}~~~~\phi_t=1
\end{equation}
depending on which are the appropriate variables.

Let us remark that the case of a fixed final condition $\rho_t$ is also of special interest, as it allows to define the quasi-potential of the process \cite{Bahadoran2010,Bouchet2016a} and gives access to its stationary measure. Moreover, if both initial and final conditions are set to be densities of fixed points or other critical manifolds of the dynamics, the optimal path connecting them is called an \textit{instanton}, and is of particular importance to estimate transition times between stationary states of multistable systems \cite{Laurie2015,Grigorio2017} or extinction times for metastable populations \cite{Assaf2010,Smith2016}.

~

For the initial condition, one possibility is to start from a fixed concentration $\rho_i$. In this case, the initial condition for the Hamiltonian trajectory is simply
\begin{equation}
\rho_0=\rho_i~~~~\mathrm{or}~~~~\phi_0=\frac{\rho_i}{\psi_0}
\end{equation}
and the density-phase variables $\{f,\rho\}$ are more appropriate, as they make the initial condition explicit.

Another popular choice is to start from a multi-Poisson distribution with average $\overline\rho$
\begin{equation}
P_0(\rho)=\prod\limits_x\frac{(V\overline\rho_x)^{V\rho_x}}{[V\rho_x]!}\mathrm{e}^{-V\overline\rho_x}~~~~\mathrm{so~that}~~~~U(\rho)=\sum\limits_x \rho_x\ln\left(\frac{\rho_x}{\overline\rho_x}\right)-\rho_x+\overline\rho_x
\end{equation}
which is to say that $U$ is a relative entropy between $\rho$ and a reference $\overline\rho$. In this case, the initial conditions become
\begin{equation}
(f_0)_x=\ln\left(\frac{\rho_x}{\overline\rho_x}\right)~~~~\mathrm{or}~~~~(\psi_0)_x=\overline\rho_x
\end{equation}
in which case the two-fields variables $\{\phi,\psi\}$ become preferable.

\subsection{Density-phase picture and global Hamiltonian attractors}
\label{IVB}

Let us first examine the behaviour of long-time trajectories in the density-phase picture:
\begin{equation}
\dot f=-\partial_\rho \mathcal{H}_{\mathfrak{s},h}~~~~\mathrm{and}~~~~\dot\rho=\partial_{f}\mathcal{H}_{\mathfrak{s},h}
\end{equation}
with the boundary conditions
\begin{equation}
\rho_0=\rho_i~~~~\mathrm{and}~~~~f_t=0.
\end{equation}

A first important remark that has to be made is that the dynamics is Hamiltonian, i.e. the volume of any subset of phase-space is conserved by the dynamics. If our boundary conditions were all at the initial or final time, that would mean that trajectories cannot converge, i.e. that no stable attractor exists, because that would require a contraction of phase-space during convergence. In other terms, all critical manifolds (fixed points, critical cycles, or stranger critical manifolds) of a Hamiltonian dynamics have either marginal or mixed stability (i.e. as many stable directions as unstable ones) and cannot be attractors if one starts from a complete initial condition $(f_i,\rho_i)$.

However, in our case, we have by construction \textit{mixed boundary conditions}: half of the boundary conditions are at the initial time, and the other half at the final time. This allows trajectories to converge towards a critical manifold of the Hamiltonian as long as the initial condition crosses its stable manifold, and the final condition crosses its unstable manifold. We will say that the critical manifold is \textit{dynamically connected} to the boundary conditions, and we will use this term in general to mean that there exists a trajectory (heterocline) from one manifold to another. This type of problems are sometimes called \textit{two-point Hamiltonian problems} in the context of optimal control \cite{Guibout2004}, although there usually both boundary conditions are for the density.

To illustrate this point, let us consider an extremely simple model with one biased reaction:
\begin{align}
\varnothing\rightarrow A~~~~&\mathrm{with~rate}~~~~a~~~~\mathrm{and~bias}~~~~~~~s\\
A\rightarrow \varnothing~~~~&\mathrm{with~rate}~~~~b~~~~\mathrm{and~bias}~~~~-s
\end{align}
The corresponding Hamiltonian is
\begin{equation}
\mathcal{H}(f,\rho)=a(\mathrm{e}^{f+s}-1)+b\rho(\mathrm{e}^{-f-s}-1).
\end{equation}
with fixed point $\{f=-s,\rho=a/b\}$. The stable manifold of this fixed point is $f=-s$, which intersects with any initial condition $\rho_i$. The unstable manifold is $\rho=\mathrm{e}^{f+s}a/b$, which intersects with the final condition $f=0$.

In the figure below, on the left, we illustrate how the trajectory solving the Hamilton equations with those boundary conditions converge to the fixed point at long times, from blue to green to red. The initial condition is the vertical purple dashed line, the final condition is the horizontal orange dashed line, the fixed point is marked in bright green, and its stable and unstable manifolds in thick black lines. Note that the time that each trajectory takes is not a simple function of its length in phase-space, but the only way to produce a long trajectory is to get closer to the fixed point, where the velocities vanish. As a result, the bulk of the trajectory converges to the fixed point, with only finite-time boundary layers connecting it to both boundary conditions. This convergence is guaranteed purely by the boundary conditions, and the critical point can be rightly called an attractor.

Similarly, we can illustrate the case of an \textit{unstable} fixed point, which occurs when the initial condition \textit{never} crosses the stable manifold, or the final condition \textit{never} crosses the unstable manifold. In that case, the long-time trajectories diverge to infinity in some direction. As we see on the figure below, on the right, if the unstable manifolds of the fixed point (red circle) are parallel to the final condition (horizontal orange dashed line), the trajectory for long times will diverge in that direction. Note however that this case is particularly unlikely to occur globally (i.e. without another stable attractor to catch the trajectory) and requires very specific symmetries of the Hamiltonian, as those manifolds have to be exactly parallel to the corresponding boundary condition. We will see an example of this in section \ref{VC}.

\begin{center}
\includegraphics[width=\textwidth]{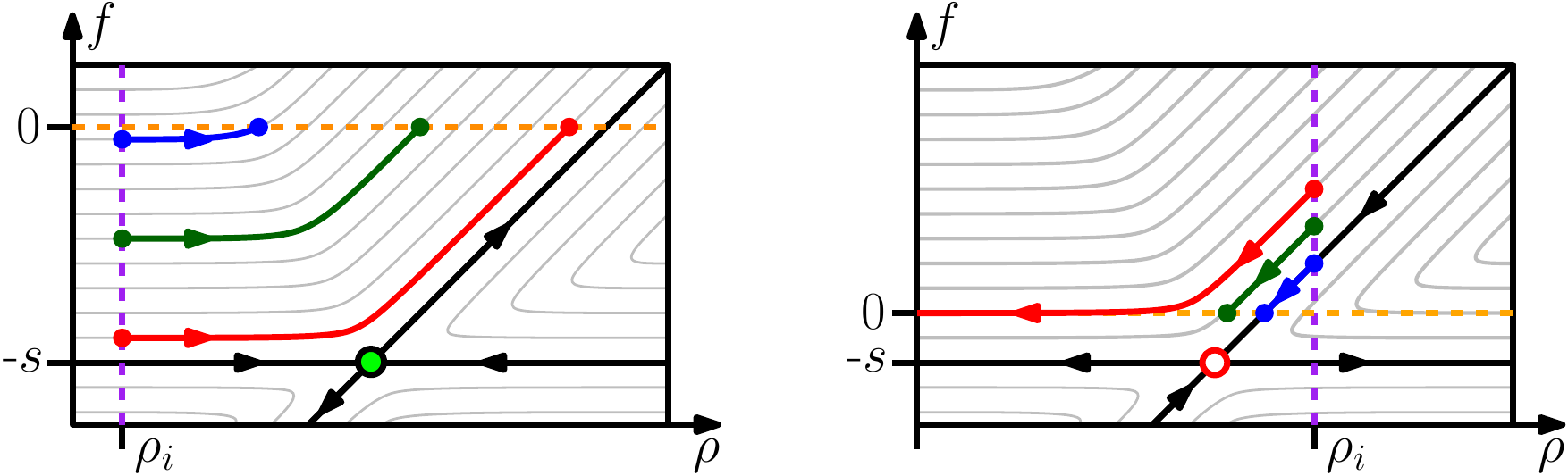}
\end{center}

We may remark that, while close to a fixed point, the kinetic term of eq.\eqref{HamSol} vanishes because $\dot\rho\sim 0$, which justifies the assumption that it is negligible. Whether it remains true for more complex attractors remains to be determined.

~~

Given these considerations, we may state the main result of this paper: the SCGF $E(\mathfrak{s},h)$ is equal to the largest value that the biased Hamiltonian $\mathcal{H}_{\mathfrak{s},h}$ takes at \textit{attractors} of the mixed-boundary conditions problem, defined as the critical manifolds that are dynamically connected to the boundary conditions, if they exist. In case the kinetic term does not vanish at all those attractors, its average value should be added to the value of the Hamiltonian.

\subsection{Conserved quantities in the two-fields picture}
\label{IVC}

We mentioned in section \ref{IIC} that chemical systems may have conservation laws: starting from a certain concentration $\rho$, the dynamics may only span a subset of phase-space (i.e. one of many ergodic components). This is not an issue in the density-phase picture, as the solution is simply to restrict the definition space of $\rho$ to the ergodic component containing the initial condition. This is not so in the two-fields picture.

~

Let us assume that the system we are considering has several conserved quantities $C^\alpha=\mathfrak{c}^\alpha\rho$, such that
\begin{equation}
{\mathrm d}_tC^\alpha={\mathrm d}_t \mathfrak{c}^\alpha\rho=-\mathfrak{c}^\alpha\nabla\cdot\lambda=0
\end{equation}
in the Lagrange frame. Let us also place ourselves in the full space of concentrations $\rho_x$, in anticipation of the fact that we will start from an initial condition that spans all space.

In the Hamiltonian frame, this has several consequences. First of all, using eq.\eqref{biasHamilton}, the previous equation expectedly becomes
\begin{equation}
\mathfrak{c}^\alpha\dot\rho=-\mathfrak{c}^\alpha\nabla\cdot\partial_{\sigma}H_{\mathfrak{s},h}(\sigma,\rho)=0~~~~\mathrm{with}~~~~\sigma=\nabla f.
\end{equation}
Moreover, when going from \eqref{biasHamilton} to \eqref{contrBiasHam}, we can no longer simply invert $\nabla$, but we have instead to consider an arbitrary integration constant which is a linear combination of the conservation laws $\mathfrak{c}^\alpha$:
\begin{equation}
\nabla\dot f=-\nabla\partial_\rho H_{\mathfrak{s},h}(\nabla f,\rho)~~~~\Rightarrow~~~~\dot f=-\partial_\rho H_{\mathfrak{s},h}(\nabla f,\rho)+A_\alpha\mathfrak{c}^\alpha
\end{equation}
where the coefficients $A_\alpha$ are entirely unconstrained and can even depend on $f$, $\rho$, or $t$. Both of those statements boil down to the fact that the Hamiltonian
\begin{equation}
\mathcal{H}_{\mathfrak{s},h}(f,\rho)=H_{\mathfrak{s},h}(\nabla f,\rho)
\end{equation}
is invariant under $f\rightarrow f+\mathfrak{c}^\alpha$ for every $\alpha$. Note that this is a gauge transformation and not a canonical change of variables: the equation of motion for $\dot f$ explicitly depends on $A_\alpha$.

This under-determination of the equations of motion is not actually an issue in the density-phase picture, since we may simply use $\nabla f$ instead of $f$ to determine the existence of attractors. Doing the same in the two-fields picture would be particularly cumbersome, since the variable $f$ has been split between $\phi$ and $\psi$. We may however solve the issue in a different way in the case where we expect our dominant attractors (those with the highest value of the Hamiltonian) to be fixed points. Assume that we know the stationary values $\rho^\star$ and $\nabla f^\star$ of $\rho$ and $\nabla f$. We then have
\begin{equation}
\nabla \dot f^\star=-\nabla\partial_\rho H_{\mathfrak{s},h}(\nabla f^\star,\rho^\star)=0
\end{equation}
which means that $\partial_\rho H_{\mathfrak{s},h}(\nabla f,\rho)$ is a linear combination of the conservation laws $\mathfrak{c}^\alpha$. We can therefore choose $A^\alpha$ uniquely such that
\begin{equation}
A_\alpha\mathfrak{c}^\alpha=\partial_\rho H_{\mathfrak{s},h}(\nabla f^\star,\rho^\star).
\end{equation}
This, in turn, implies that 
\begin{equation}
\dot f=-\partial_\rho H_{\mathfrak{s},h}(\nabla f,\rho)+A_\alpha\mathfrak{c}^\alpha=0~~~~\mathrm{when}~~~~\rho=\rho^\star~~,~~\nabla f=\nabla f^\star
\end{equation}
which means that any compatible value of $\{f,\rho\}$ is a fixed point. Note that the previous equation is also a way to determine $A_\alpha$ if they are considered as unknowns.

The final step is to translate those equations in the two-field picture. Since the only difference is in $\dot f$, considering the derivation in section \ref{IIIA4}, we simply get
\begin{equation}
\dot\psi_x=\partial_{\phi_x}\mathscr{H}_{\mathfrak{s},h}-A_\alpha\mathfrak{c}^\alpha_x\psi_x=0~~~~\mathrm{and}~~~~\dot\phi_x=-\frac{1}{\psi_x}\partial_{\phi_x}\mathscr{H}_{\mathfrak{s},h}+A_\alpha\mathfrak{c}^\alpha_x\phi_x=0.
\end{equation}

~~

In conclusion, we arrive at a second formulation of our main result, in the case where all attractors are fixed points: the SCGF $E(\mathfrak{s},h)$ is equal to the largest value that the biased Hamiltonian $\mathcal{H}_{\mathfrak{s},h}$ takes at solutions of
\begin{equation}\boxed{\boxed{
\frac{1}{\psi_x}\partial_{\phi_x}\mathscr{H}_{\mathfrak{s},h}=\frac{1}{\phi_x}\partial_{\psi_x}\mathscr{H}_{\mathfrak{s},h}=-A_\alpha\mathfrak{c}^\alpha
}}
\end{equation}
where the factors $A_\alpha$ are unknowns of the equations.

Note that, in the special case of a linear chemical network, where every complex is a different monomer (or a different conformation of a single molecule), and where the only conserved quantity is the total mass (i.e. $\mathfrak{c}=1$), those equations translate precisely as eigenvalue equations for $K_{s,h}$, and we recover the well-known result mentioned in section \ref{IID}.

\subsection{Generic dynamical phase transitions for multistable systems}
\label{IVD}

One of the interesting features of this last result is the following: if the dynamics has several fixed points with different values of the Hamiltonian, whenever the two largest values cross each-other due to varying the parameters $\mathfrak{s}$ and $h$, the system undergoes a first order dynamical phase transition. In this section, we will see that this scenario is almost guaranteed in systems where the deterministic (unbiased) dynamics has several fixed points.

~

We recall the expression of the contracted biased Hamiltonian:
\begin{equation}
\mathcal{H}_{\mathfrak{s},h}(f,\rho)=\sum\limits_{\gamma,\gamma'} k_{\gamma'\gamma}\rho^{\nu^\gamma}\left(\mathrm{e}^{(\nu^{\gamma'}-\nu^\gamma)f+\mathfrak{s}_{\gamma'\gamma}}-1\right)+\sum\limits_x h_x \rho_x.
\end{equation}
with dynamics
\begin{equation}\label{pertHam}
\dot f=-\partial_\rho \mathcal{H}_{\mathfrak{s},h}~~~~\mathrm{and}~~~~\dot\rho=\partial_{f}\mathcal{H}_{\mathfrak{s},h},
\end{equation}
and boundary conditions
\begin{equation}
\rho_0=\rho_i~~~~\mathrm{and}~~~~f_t=0.
\end{equation}

The unbiased version $\mathfrak{s}=h=0$ takes a particularly simple form: it is straightforward to check that $\dot f=0$ whenever $f=0$, which implies that $f=0$ along any solution due to the final condition. The equation on $\rho$ then becomes
\begin{equation}
\dot\rho_x=\partial_{f}\mathcal{H}_{0,0}(0,\rho)=\sum\limits_{\gamma,\gamma'} (\nu_x^{\gamma'}-\nu_x^\gamma)k_{\gamma'\gamma}\rho^{\nu^\gamma}
\end{equation} 
which is the usual deterministic mass-action dynamics.

Consider the case where this equation admits several fixed points $\dot\rho^{(0)}_i=0$. The value of the unbiased Hamiltonian is $0$ along the whole $f=0$ space, which means it is $0$ at each of those fixed points, and since they are the only ones dynamically connected to the final condition, they are all potential convergence points, depending on the initial condition.

We now turn on the bias, with a global infinitesimal variable $\varepsilon$, and we look for the positions of the new fixed points to first non-zero order in $\varepsilon$:
\begin{align}
\rho_i(\varepsilon)&\sim\rho^{(0)}_i+\varepsilon \rho^{(1)}_i+\frac{\varepsilon^2}{2} \rho^{(2)}_i+\dots\\
f_i(\varepsilon)&\sim 0+\varepsilon f^{(1)}_i+\frac{\varepsilon^2}{2} f^{(2)}_i+\dots\\
\mathcal{H}_\varepsilon(f,\rho)&\sim \mathcal{H}_{0,0}(f,\rho)+\varepsilon \mathcal{H}^{(1)}(f,\rho)+\frac{\varepsilon^2}{2} \mathcal{H}^{(2)}(f,\rho)+\dots.
\end{align}
Equations \eqref{pertHam} give us, to first order (order zero trivially vanishes):
\begin{align}
\dot\rho^{(1)}_i&=\partial_{f}\mathcal{H}^{(1)}(0,\rho^{(0)}_i)+\partial^2_{f}\mathcal{H}_{0,0}(0,\rho^{(0)}_i)f^{(1)}_i+\partial_{\rho}\partial_{f}\mathcal{H}_{0,0}(0,\rho^{(0)}_i)\rho^{(1)}_i=0\\
\dot f^{(1)}_i&=-\partial_{\rho}\mathcal{H}^{(1)}(0,\rho^{(0)}_i)-\partial_{f}\partial_\rho\mathcal{H}_{0,0}(0,\rho^{(0)}_i)f^{(1)}_i-\partial^2_{\rho}\mathcal{H}_{0,0}(0,\rho^{(0)}_i)\rho^{(1)}_i=0
\end{align}
which is a linear equation on $\{f^{(1)}_i,\rho^{(1)}_i\}$ that always has a solution if the Wronskian $W_i=\partial^2 \mathcal{H}_{0,0}(0,\rho^{(0)}_i)$ (i.e. the matrix of second derivatives) is invertible, which is generally the case:
\begin{equation}
\begin{bmatrix}f^{(1)}_i\\\rho^{(1)}_i\end{bmatrix}=-W^{-1}_i\cdot \begin{bmatrix}\partial_{f}\mathcal{H}^{(1)}(0,\rho^{(0)}_i)\\\partial_{\rho}\mathcal{H}^{(1)}(0,\rho^{(0)}_i)\end{bmatrix}.
\end{equation}
The new value of the Hamiltonian is then
\begin{equation}
\mathcal{H}_\varepsilon(f_i,\rho_i)\sim \mathcal{H}_{0,0}(0,\rho^{(0)}_i)+\varepsilon\left(\partial_{f}\mathcal{H}_{0,0}(0,\rho^{(0)}_i)f^{(1)}_i+\partial_{\rho}\mathcal{H}_{0,0}(0,\rho^{(0)}_i)\rho^{(1)}_i+ \mathcal{H}^{(1)}(0,\rho^{(0)}_i)\right)
\end{equation}
which is to say
\begin{equation}\boxed{\boxed{
\mathcal{H}_\varepsilon(f,\rho)\sim\varepsilon\mathcal{H}^{(1)}(0,\rho^{(0)}_i).
}}
\end{equation}
This expression does not depend on the specific values of $\{f^{(1)}_i,\rho^{(1)}_i\}$ but only on their existence and the value of the perturbation at the original fixed point. This should come as no surprise to anyone familiar with perturbation theory for eigenvalues and eigenstates of matrices: the first order in perturbation of the eigenvalues only depend on the value of the perturbation in the original eigenstates.

~

For generic values of biases, the value of that perturbation will be different at every fixed point, which means that they all become dynamically disconnected (since all trajectories conserve the value of the Hamiltonian). Moreover, the fixed point where that value is highest will generically depend on the direction of the perturbation, which means that the dominant attractor of our dynamics will jump from one fixed point to another when crossing $\varepsilon=0$. This allows us to conclude that multistable chemical systems generically undergo first-order dynamical phase transitions at zero bias. We will see examples of this in simple cases in section \ref{V}. In particular, we will exhibit examples where the dominant stable fixed points of the biased dynamics are \textit{unstable} fixed points in the unbiased system, which get stabilised by fluctuations.

Note that this result requires $\mathcal{H}^{(1)}(0,\rho)$ to not be independent of $\rho$, which is the case for most nontrivial biases. Moreover, the argument above can just as easily be made for more complex attractors, such as limit cycles. However, the potential contribution of the kinetic term $f\dot\rho$ in the extremisation problem makes it less clear which attractor will dominate for small biases. We therefore leave this case as a conjecture for the time being.

\section{A few examples of dynamical phase transitions}
\label{V}

In this final section, we will illustrate our results with various examples of dynamical phase transition that we may observe in multistable stochastic chemical reaction networks. We will consider the Schl\"ogl model \cite{Schlogl1972}, which is one of the simplest bistable chemical models, as well as variants of it with different numbers of fixed points, and we will see how those fixed points undergo phase transitions as we vary the bias. Two surprising observations will be made: fluctuations can restore broken ergodicity in the large volume limit, and can stabilise unstable fixed points of the deterministic dynamics.

Even though all previous formulae apply to both density and current biases, we will only consider the latter, i.e. we take $h=0$, for the sake of simplicity. An example of a dynamical phase transition involving a density bias can be found in \cite{Zilber2019}.

\subsection{Schl\"ogl model}
\label{VA}

In the Schl\"ogl model, two reversible chemical reactions can occur, with the following kinetic rates:
\begin{align}
~\varnothing &\rightarrow ~A~~~~~\mathrm{with~rate}~~~~~k_0~~~~~\mathrm{and~bias}~~~~~s_0\nonumber\\
~A &\rightarrow ~\varnothing~~~~~\mathrm{with~rate}~~~~~k_1~~~~~\mathrm{and~bias}~~~~~s_1\nonumber\\
2A &\rightarrow 3A~~~~~\mathrm{with~rate}~~~~~k_2~~~~~\mathrm{and~bias}~~~~~s_2\nonumber\\
3A &\rightarrow 2A~~~~~\mathrm{with~rate}~~~~~k_3~~~~~\mathrm{and~bias}~~~~~s_3\nonumber.
\end{align}

The biased Hamiltonian of the model is given by
\begin{equation}
\mathcal{H}_{\mathfrak{s},0}(f,\rho)=k_0(\mathrm{e}^{f+s_0}-1)+k_1\rho(\mathrm{e}^{-f+s_1}-1)+k_2\rho^2(\mathrm{e}^{f+s_2}-1)+k_3\rho^3(\mathrm{e}^{-f+s_3}-1)
\end{equation}
and the deterministic equation is
\begin{equation}
\dot\rho=\partial_f\mathcal{H}_{0,0}(f,\rho)=k_0-k_1\rho+k_2\rho^2-k_3\rho^3.
\end{equation}
We start from a fixed initial condition $\rho_0$.

For an appropriate choice of rates (e.g. $k_0=0.7$, $k_1=2$, $k_2=0.9$ and $k_3=0.1$), this equation can have three fixed points, of which the middle one is unstable, and the others stable. On the figure below, we represent the Hamiltonian trajectories for the unbiased system, where the stable fixed points are represented in green, and the unstable one in red. The sign of the Hamiltonian is indicated in each region, as a reference for the following deformations, in order to identify the critical point with highest value of $\mathcal{H}$. Note that, in this context, the whole dynamics takes place on the $f=0$ line due to the final condition, and that the notion of stability of fixed points is therefore different from the fluctuating case. Also note that the stable fixed point that will be reached depends on the initial condition, which is due to the breaking of ergodicity in the large-volume limit \cite{Oppenheim1977,Vroylandt2018}.

\begin{center}
\includegraphics[width=0.4\textwidth]{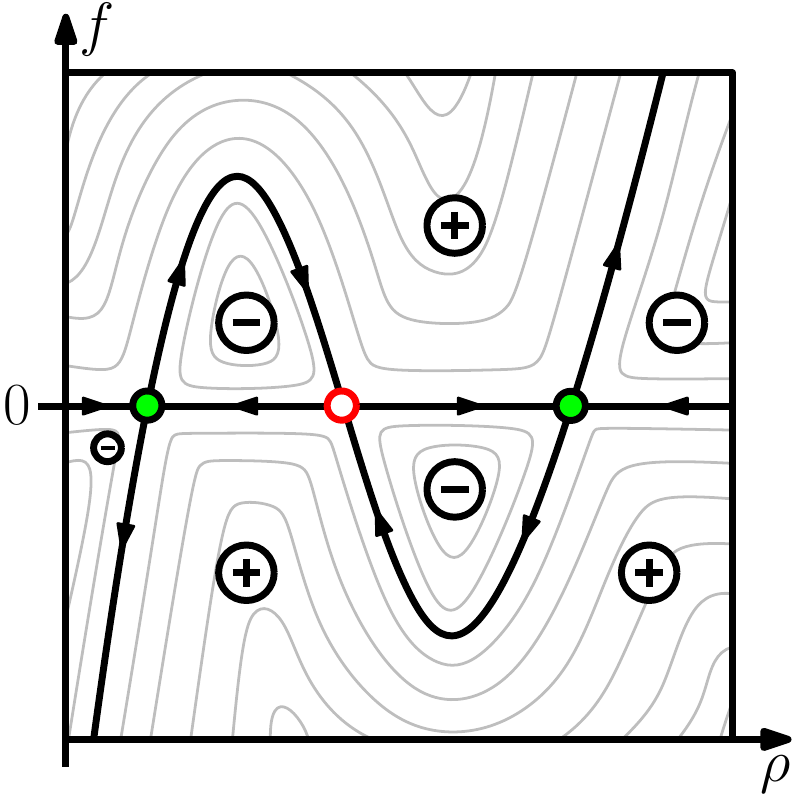}
\end{center}

We now turn on the bias, for example by conditioning on the current of the reaction $\varnothing\leftrightarrow A$ alone: $s_2=-s_3=s$ and $s_0=s_1=0$. The new trajectories are shown on the following figure, depending on the sign of $s$.

\begin{center}
\includegraphics[width=\textwidth]{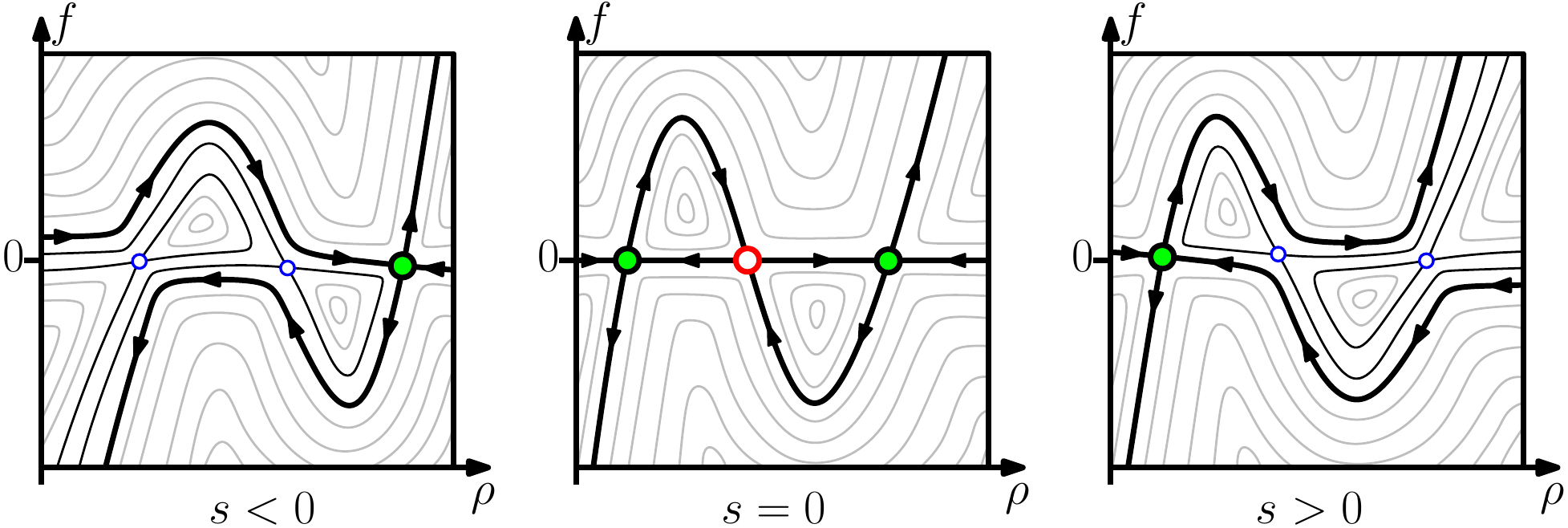}
\end{center}

We represent the fixed point with the highest value of the Hamiltonian in green. As we see, it jumps from right to left when $s$ crosses $0$, and in both cases this fixed point is unique regardless of the initial condition, as the other two are exponentially subdominant. This can be easily understood when looking at the value of the perturbation at first order in $s$ across the $f=0$ line:
\begin{equation}\label{sH1}
s\mathcal{H}^{(1)}(0,\rho)=k_0s_0+k_1\rho s_1+k_2\rho^2s_2+k_3\rho^3s_3=s(k_0-k_1\rho),
\end{equation}
which is increasing if $s<0$ and decreasing if $s>0$, consistently with which fixed point is favoured. On the figure below, on the left, we plot the values of $\mathcal{H}$ on at all three fixed points as a function of $s$. The red curve corresponds to the lowest density, the blue one to the highest, and the green one to the central fixed point. As we see, for a positive $s$ the blue curve is highest, i.e. the rightmost point is dominant, whereas for a slightly negative $s$ the red curve (leftmost point) is dominant. The SCGF $E(s)$ is obtained by keeping the maximum of all three curves, and has an non-analyticity at $s=0$. On the right figure, we draw the resulting long-time large deviation function $g(\lambda)$ obtained by computing the Legendre transform of $E(s)$. We see that it exhibits a horizontal plateau, approximately between $\lambda=0.2$ (as visible on the inset) and $\lambda=12$, which is an expected signature of a first-order phase transition at $s=0$ (the slope of the plateau is consistent with the value of $s$). The sharp increase in slope for $\lambda<0$ is due to the fluctuation relation, and is smooth, as can be seen on the inset, so that there is no phase transition at $\lambda=0$ even though the figure might suggest it.

\begin{center}
\begin{minipage}{0.45\textwidth}
\includegraphics[width=\textwidth]{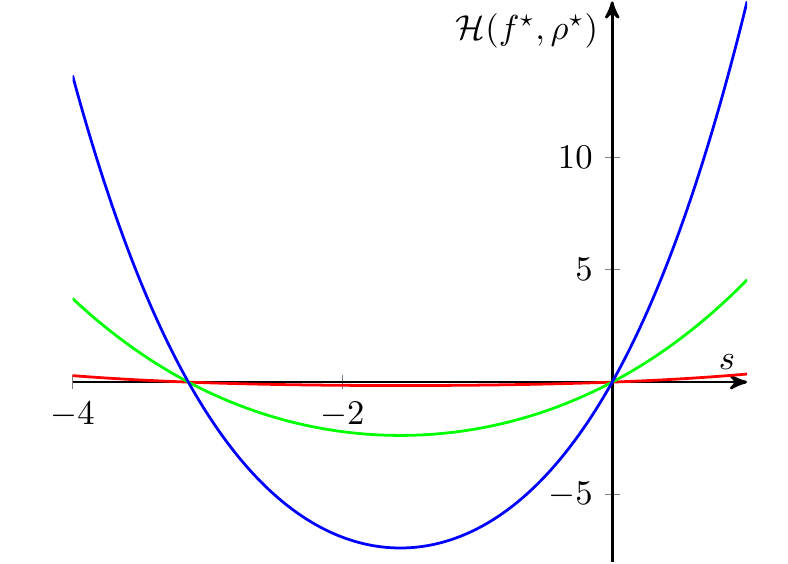}
 \end{minipage}
 ~~~~~~~
\begin{minipage}{0.45\textwidth}
\includegraphics[width=\textwidth]{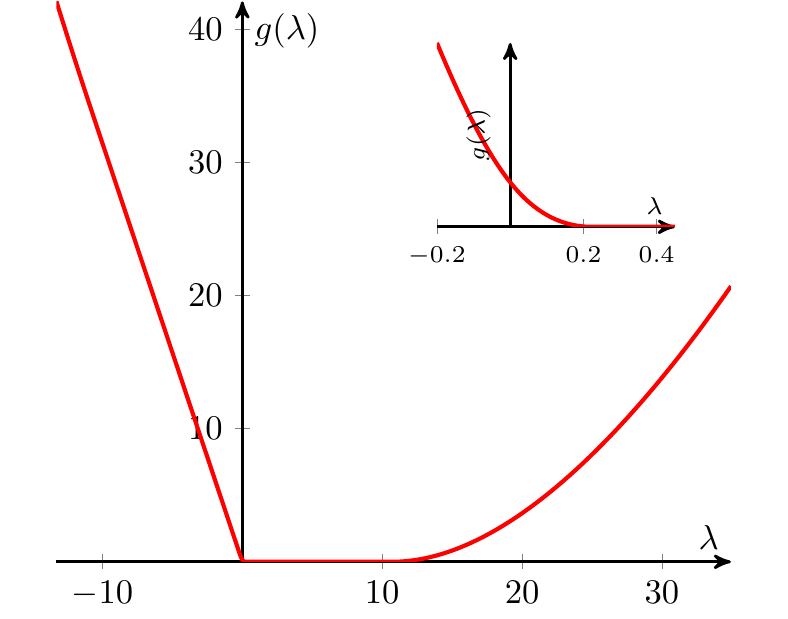}
 \end{minipage}
\end{center}

~~

~~

A surprising observation is that, on both sides of the transition, the ergodicity of the microscopic system, lost in the large-volume limit, is restored: any initial condition leads to the same fixed point, as its stable manifold spans all density-space.

~

In light of eq.\eqref{sH1}, we may wonder if it is possible to favour the central fixed point instead. It is indeed the case, as it is quite easy to engineer a bias which makes the perturbation maximal around that value of $\rho$. Consider for instance $s_0=-4s$, $s_1=3s$, $s_2=-s$ and $s_3=0$ for $s$ small and positive. The resulting trajectories are represented on the following figure.

\begin{center}
\includegraphics[width=0.4\textwidth]{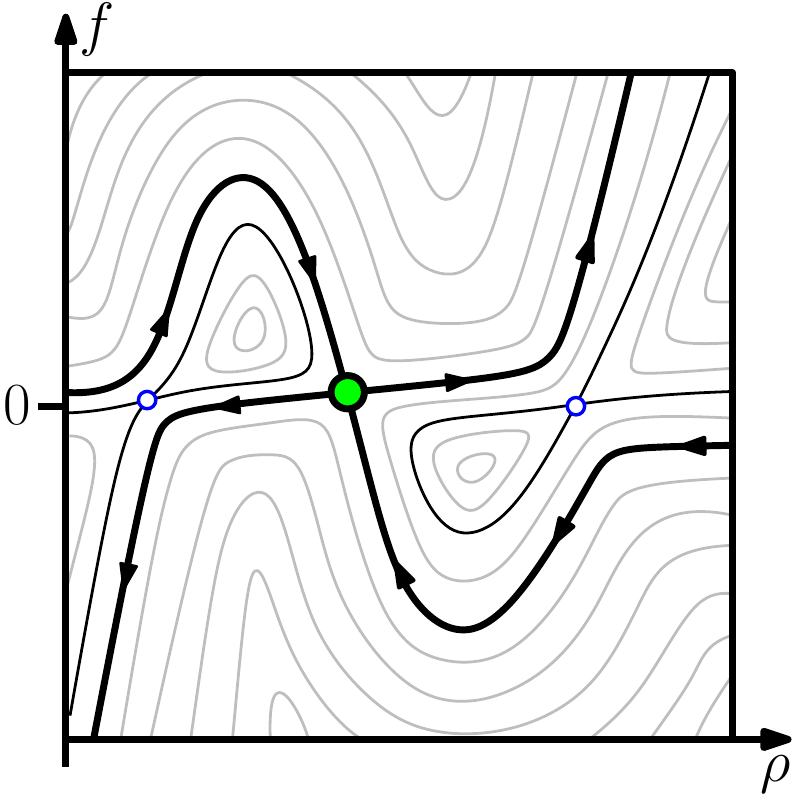}
\end{center}

The conclusion may seem quite counterintuitive: there is no fundamental difference between stable and unstable deterministic attractors when looking at biased trajectories, and fluctuations can stabilise unstable stationary states, just as an extra external driving can \cite{Basu2016}.

\subsection{Generalised Schl\"ogl model}
\label{VB}

It is straightforward to generalise the previous case to one with any number of fixed points, by considering reactions of any order:
\begin{align}
nA &\rightarrow (n+1)A~~~~~\mathrm{with~rate}~~~~~k^+_{n}~~~~~\mathrm{and~bias}~~~~~s^+_{n}\nonumber\\
nA &\rightarrow (n-1)A~~~~~\mathrm{with~rate}~~~~~k^-_{n}~~~~~\mathrm{and~bias}~~~~~s^-_{n}\nonumber
\end{align}
and choosing the rates such that the deterministic equation
\begin{equation}
\dot\rho=\sum\limits_{n=n_{\mathrm{min}}}^{n_{\mathrm{max}}} (k^+_n-k^-_n)\rho^n
\end{equation}
has the desired number of fixed points. Any of the fixed points can then become dominant under the appropriate perturbation
\begin{equation}
s\mathcal{H}^{(1)}(0,\rho)=\sum\limits_{n=n_{\mathrm{min}}}^{n_{\mathrm{max}}} (k^+_{n}s^+_{n}+k^-_{n}s^-_{n})\rho^n.
\end{equation}

Note that as long as the lowest reaction is creative ($k^-_{n_{\mathrm{min}}}=0$) and the highest one is destructive ($k^+_{n_{\mathrm{max}}}=0$), the first and last fixed points will be stable in the deterministic case. This also implies that the non-zero solution of $\mathcal{H}_{0,0}(f,\rho)=0$ spans the whole $f$-space: we have
\begin{equation}
\mathcal{H}_{0,0}(f,\rho)=\sum\limits_{n=n_{\mathrm{min}}}^{n_{\mathrm{max}}} (\mathrm{e}^f-1)k^+_{n}\rho^n+ (\mathrm{e}^{-f}-1)k^-_{n}\rho^n=(\mathrm{e}^f-1)\sum\limits_{n=n_{\mathrm{min}}}^{n_{\mathrm{max}}} \left(k^+_{n}- \mathrm{e}^{-f}k^-_{n}\right)\rho^n
\end{equation}
which cancels at $f=0$ or at
\begin{equation}
f^\star(\rho)=\ln\left(\frac{\sum_n k^-_{n}\rho^n}{\sum_n k^+_{n}\rho^n}\right)
\end{equation}
such that
\begin{equation}
f^\star(\rho\rightarrow 0)\rightarrow-\infty~~~~\mathrm{and}~~~~f^\star(\rho\rightarrow \infty)\rightarrow+\infty.
\end{equation}
This, in turn, implies that there will always be at least one fixed point dynamically connected to the boundary conditions. In the following example, we see what may happen when this is not the case.

\subsection{Runaway Schl\"ogl model}
\label{VC}

As a final example, we consider a variant of the Schl\"ogl model whose deterministic dynamics is not always stable: the so-called \textit{runaway} Schl\"ogl model \cite{Elgart2004}. It is obtained by taking $k_3=0$ in the Schl\"ogl model, so that the stability condition mentioned above is not true anymore: the highest-order reaction is creative, and any initial condition larger that the largest fixed point will diverge to infinity.

Let us now consider a bias $s_0=s+m$, $s_1=-s-m$ and $s_2=m$. This lets us effectively change the final condition through $m$ and favour one or the other fixed point through $s$. On the figure below, we draw the Hamiltonian trajectories of the model for $m=0$ and $s<0$, $s=0$ and $s>0$.

\begin{center}
\includegraphics[width=\textwidth]{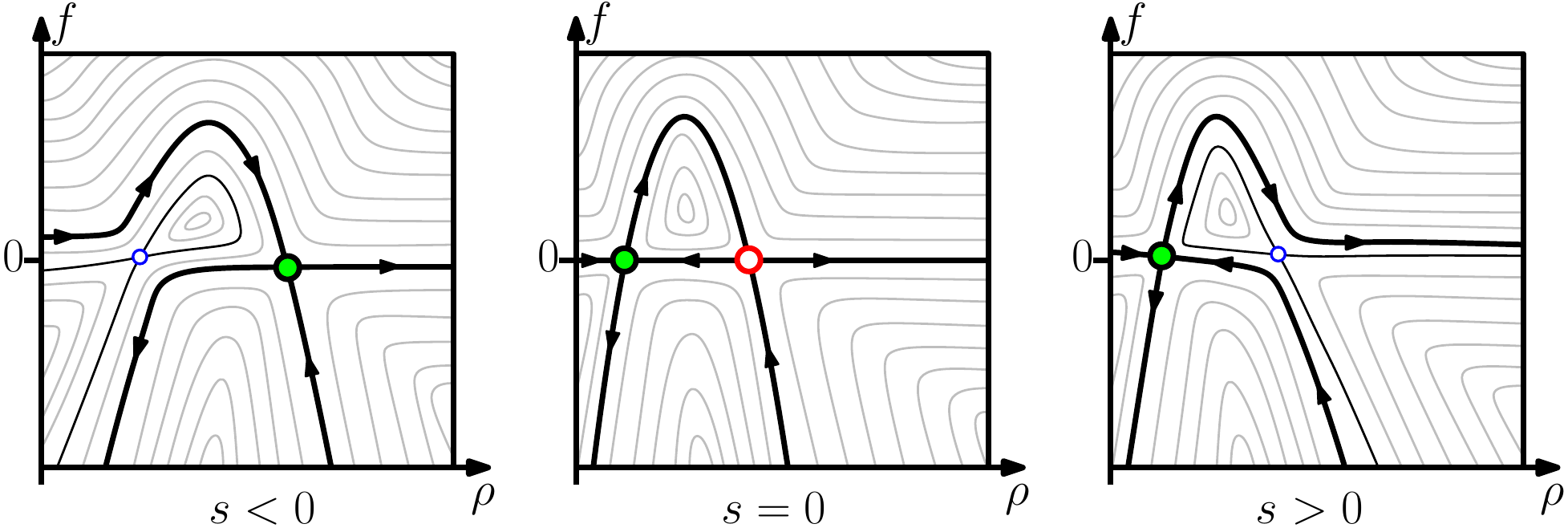}
\end{center}

Once again, we find a first-order dynamical phase transition between the two fixed points, one of them being originally unstable. Moreover, we see that in all cases, the unstable manifolds of the fixed points cannot reach beyond a certain value $f_{\mathrm{max}}$ of $f$. If we set $m$ to be above that critical value, the fixed point becomes dynamically disconnected from the boundary conditions, and the system becomes unstable. This is shown on the following figure, where the initial condition is shown in purple (vertical dashed line), and the final conditions in orange (horizontal dashed lines, for two different values of $m$). For a final condition below the dominant fixed point, there is a stable trajectory converging to it, shown in green. However, for a final condition above $f_{\mathrm{max}}$, the only trajectories that have the right length are those shown in red, and diverge as time grows. If the final condition is between those two regions, both types of trajectories exist, and it is unclear which dominates, as the red trajectories have a diverging value of the Hamiltonian but a nonzero kinetic term $f\dot\rho$.

\begin{center}
\includegraphics[width=0.4\textwidth]{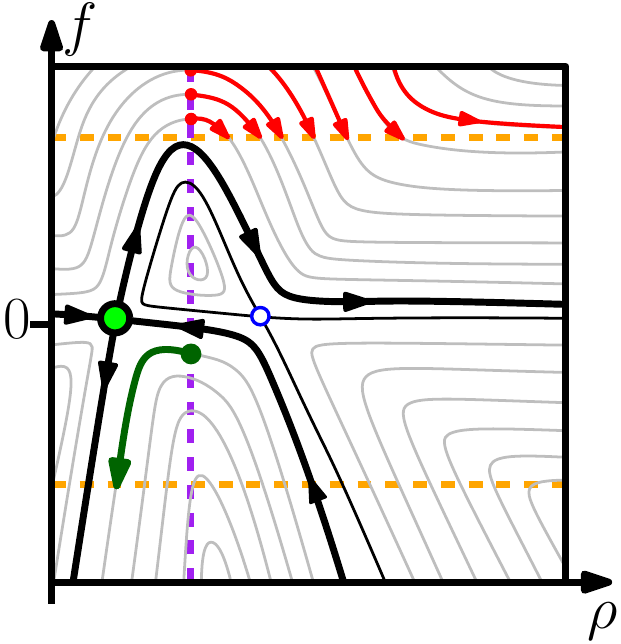}
\end{center}

This example gives us yet another type of dynamical phase transitions that can be found in some stochastic chemical networks, specifically ones that are deterministically unstable. A similar phase transition was observed in \cite{Nickelsen2018} in an Orstein-Uhlenbeck process.

\section{Conclusion}
\label{VI}

We have proven that, in the limit of large volumes, a chemical master equation for a well-mixed chemical reaction network can be described through a large-deviations path integral over a hydrodynamic Lagrangian. We have derived that Lagrangian explicitly from the microscopic dynamics by keeping track of every individual chemical current. We have then derived the equations that the minimiser of said path integral should verify, either through Euler-Lagrange equations or through the definition of an appropriate Hamiltonian and the corresponding Hamilton equations. We have shown that this formalism extends naturally to the case where the dynamics is biased with respect to time-additive observables, and that the scaled cumulant generating function of those observables can be expressed in terms of the largest value that the biased Hamiltonian takes over all solutions at fixed time. In the specific case of mass-action dynamics, a second set of canonical coordinates is available, leading to an equivalent of the Doi-Peliti Hamiltonian which is appropriate for dealing with conserved quantities.

The boundary conditions for those equations, which turn out to be split between the initial and final coordinates, are of crucial importance, as they allow the solutions to converge to global attractors even though the dynamics is Hamiltonian. The aforementioned cumulant generating function can then be identified as the maximal value the Hamiltonian takes across those attractors, provided that the kinetic term of the action is negligible. In the case where the deterministic chemical reaction network has several fixed points, we have shown that, under bias, the system generically undergoes first-order dynamical phase transitions around zero bias, which we have illustrated through variants of the Schl\"ogl model. Finally, we have shown an example of a dynamical phase transition where the system becomes unstable on one side of the transition.

~~

Here are a few important takeaways from both our abstract results and illustrative examples:
\begin{itemize}
\item \textbf{tracking more is easier}: the detailed Lagrangian, which keeps track of all individual chemical currents, is explicit, whereas the standard Lagrangian is usually not;
\item \textbf{slicing with precision is crucial}: taking a time-step $\delta t$ small enough to vanish but large enough that $V\delta t$ diverges is essential to obtain the correct Lagrangians;
\item \textbf{boundary conditions make all the difference}: a Hamiltonian dynamics with a full initial condition cannot have attractors, but one with mixed boundary conditions often does;
\item \textbf{stability is relative}: a fixed point which is unstable in the deterministic dynamics of a system may become stable under even an infinitesimal bias;
\item \textbf{what is broken can be fixed}: a system with microscopic ergodicity, broken due to some scaling limit, might well see it restored when constrained in certain ways.
\end{itemize}

~

There are many problems still open on the topic of large deviations of population models. Chief amongst them is the case of systems with attractors such as limit cycles or even strange attractors, for which very little it known on the large-deviations scale, beyond some numerical results on entropy production \cite{Gaspard2004}. In \cite{Smith2016}, the probability of extinction in a model with a cycle and a fixed point at zero population was estimated, by finding the instanton connecting the cycle to the fixed point through a Hamiltonian formalism. Perhaps similar methods can be applied for models biased on currents or densities, although even the shape of the biased attractors is yet to be determined.

We have also avoided the case of systems sitting at the critical point of a stationary phase transition, where any bias might force the system into one phase or the other. The Schl\"ogl model is simple enough to treat numerically at its critical point, but whether those critical points are special in terms of which dynamical phase transitions can occur under bias remains to be seen.

A crucial concept in chemical networks, that is only indirectly relevant to this work, is that of deficiency \cite{Polettini2015}: in essence, it is the number of conservation laws that are not visible immediately on the network of complexes. That number has important consequences on the types of stationary states that one may observe \cite{Feinberg1987}. If it is $0$, for instance, it is known that the system has exactly one stable fixed point, and no other attractor \cite{Anderson2010a}. It would be interesting to determine the consequences of deficiency in the context of dynamical large deviations.

We have mentioned the delicate case of dynamics which come close to the boundaries of concentration space, where the number of individuals of one or more species comes back down to a microscopic scale. Those systems can be treated as so-called \textit{hybrid models}, where part of the dynamics is approximated though a large deviations formalism and the rest is kept as a Markov jump process \cite{Winkelmann2017,Andreis2019}. This would apply for instance to systems with a small number of enzymes who go through various conformations \cite{Li2016}. A few results exist as regards the large deviations of such systems \cite{Bressloff2014}, but as far as we know the probability distributions of time-averaged observables have not been tackled so far.

Finally, one may naturally wonder how the structure and phenomenology of those models would be affected if one adds spatial coordinates and lets the chemicals diffuse, as in reaction-diffusion models \cite{Jona-Lasinio1993}. It is known that for instance, in certain simple 1-species models with an active/inactive stationary transition (i.e. where, depending on a rate constant, the system has a fixed point at a nonzero/zero density), the position of the transition is strongly affected by the dimension of the diffusion space, to the point where special nonperturbative renormalisation techniques \cite{Canet2011} are required to account for it properly in $D\geq 3$ \cite{Canet2004,Gredat2014}. Whether anything similar occurs for multistable models, and how this interacts with dynamical biases and phase transitions, shall be a very interesting question for a future work. We might also ask similar questions for systems that exhibit instabilities leading to Turing patterns \cite{Falasco2018}.

~

The authors would like to thank Riccardo Rao, Artur Wachtel, Matteo Polettini, Gatien Verley, Hadrien Vroylandt, Christian Maes and Richard Kraaij for many useful discussions.

\newpage

\section{Appendix}
\label{Appendix}

In this appendix, we will compute explicitly the microscopic transition probabilities of our chemical networks, with or without bias, in the limit of a large volume and a short time-step: $V\rightarrow\infty$, $\delta t\rightarrow 0$. This will be done by exploiting the fact that, in those limits, all separate reactions essentially commute.

Consider the microscopic chemical Markov matrix
\begin{equation}
W=\sum\limits_{\gamma', \gamma,n}W_{\gamma'\gamma}(n)~| n-\nabla_{\gamma', \gamma}\rangle\langle n|-W_{\gamma'\gamma}(n)~| n\rangle\langle n|.
\end{equation}
The transition probability from $n_i$ to $n_f$ during time $\delta t$ is
\begin{equation}
\mathrm{P}_{\delta t}[n_f|n_i]=\langle n_f|\mathrm{e}^{\delta tW}|n_i\rangle.
\end{equation}
Moreover, the generating function of exponential moments of the observables $\lambda$ and $\rho$ defined in section \ref{IIB} between those states and during that time, with respective conjugate variables $\mathfrak{s}$ and $h$, is given by \cite{Touchette20091,Lazarescu2015}:
\begin{equation}
\langle \mathrm{e}^{\delta t( \mathfrak{s}\cdot\lambda+h\rho)}\rangle^{n_f}_{n_i}=\langle n_f|\mathrm{e}^{\delta tW_{\mathfrak{s},h}}|n_i\rangle.
\end{equation}
where the Markov matrix has been replaced by its biased version
\begin{equation}
W_{\mathfrak{s},h}=\sum\limits_{\gamma', \gamma,n}\mathrm{e}^{\mathfrak{s}_{\gamma' \gamma}}W_{\gamma'\gamma}(n)~| n-\nabla_{\gamma', \gamma}\rangle\langle n|-W_{\gamma'\gamma}(n)~| n\rangle\langle n|+\sum\limits_{x,n} h_x n_x~| n\rangle\langle n|.
\end{equation}
The effect of this biasing on the probability of each trajectory is to multiply it by $\mathrm{e}^{\mathfrak{s}_{\gamma' \gamma}}$ for each occurrence of a reaction $\gamma\rightarrow\gamma'$ regardless of the starting composition $n$, thus generating the moments of $\lambda$, and by $\mathrm{e}^{\int\!hn(\tau)\mathrm{d}\tau}$ which generates moments of $\rho$.

~~

We will now simplify this expression in the limit of a large volume $V$, in the case of mass-action kinetics
\begin{equation}
W_{\gamma'\gamma}(n)=k_{\gamma'\gamma}\prod_{x}\frac{[n_x]!}{[n_x-\nu_x^\gamma]!}V^{1-\sum_x \nu_x^\gamma},
\end{equation}
by showing that all individual reaction operators
\begin{equation}
A^{\gamma'\gamma}_{\mathfrak{s}}=\sum\limits_n\mathrm{e}^{\mathfrak{s}_{\gamma' \gamma}}W_{\gamma'\gamma}(n)~| n-\nabla_{\gamma', \gamma}\rangle\langle n|
\end{equation}
as well as the corresponding escape rates
\begin{equation}
B^{\gamma'\gamma}=-\sum\limits_n W_{\gamma'\gamma}(n)~| n\rangle\langle n|
\end{equation}
and the concentration-counting operator
\begin{equation}
B_{h}=\sum\limits_{x,n} h_x n_x~| n\rangle\langle n|.
\end{equation}
all commute on the large deviations scale for short times, as long as each particle number $n$ is of order $V$.

Let us first consider $A_1=A^{\gamma'\gamma}_{\mathfrak{s}}$, $A_2=A^{\mu'\mu}_{\mathfrak{s}}$ and $B=B^{\gamma'\gamma}$ and compute their commutator:
\begin{align}
A_2& A_1-A_1 A_2=\nonumber\\
&~~~\sum\limits_n \mathrm{e}^{\mathfrak{s}_{\gamma' \gamma}+\mathfrak{s}_{\mu' \mu}}\Bigl(W_{\mu'\mu}(n-\nabla_{\gamma', \gamma})W_{\gamma'\gamma}(n)-W_{\gamma'\gamma}(n-\nabla_{\mu', \mu})W_{\mu'\mu}(n) \Bigr)|n-\nabla_{\gamma', \gamma}-\nabla_{\mu', \mu}\rangle\langle n|\nonumber\\
A_2&B-B A_2=\nonumber\\
&-\sum\limits_n\mathrm{e}^{\mathfrak{s}_{\mu' \mu}}\Bigl(W_{\mu'\mu}(n)W_{\gamma'\gamma}(n)-W_{\gamma'\gamma}(n-\nabla_{\mu', \mu})W_{\mu'\mu}(n) \Bigr) |n-\nabla_{\mu', \mu}\rangle\langle n| \nonumber.
\end{align}
For $n$ of order $V$, we can expand $W_{\gamma'\gamma}(n)$ in orders of $V$ using Stirling's approximation, defining $\rho=n/V$:
\begin{align}
W_{\gamma'\gamma}(n)&\sim k_{\gamma'\gamma}V\prod_y \rho_y^{\nu^\gamma_y}\left(1+\sum_x\frac{\nu^\gamma_x(1-2\nu^\gamma_x)}{2V\rho_x} \right)\nonumber\\
W_{\gamma'\gamma}(n-\nabla_{\mu', \mu})&\sim k_{\gamma'\gamma}V\prod_y \rho_y^{\nu^\gamma_y}\left(1+\sum_x\frac{\nu^\gamma_x(1-2\nu^\gamma_x)}{2V\rho_x} -\nu^\gamma_x\frac{\nu^{\mu'}_x-\nu^\mu_x}{V\rho_x}\right)\nonumber
\end{align}
so that
\begin{align}
A_2& A_1-A_1 A_2=\nonumber\\
&~~~V\sum\limits_n \mathrm{e}^{\mathfrak{s}_{\gamma' \gamma}+\mathfrak{s}_{\mu' \mu}}~k_{\mu'\mu}k_{\gamma'\gamma}\rho^{\nu^\mu+\nu^\gamma}\left(\sum_x\frac{\nu_x^\gamma\nu_x^{\mu'}-\nu_x^\mu\nu_x^{\gamma'}}{\rho_x}\right)|n-\nabla_{\gamma', \gamma}-\nabla_{\mu', \mu}\rangle\langle n|\nonumber\\
A_2&B-B A_2=\nonumber\\
&-V\sum\limits_n\mathrm{e}^{\mathfrak{s}_{\mu' \mu}}~k_{\mu'\mu}k_{\gamma'\gamma}\rho^{\nu^\mu+\nu^\gamma}\left(\sum_x\nu^\gamma_x\frac{\nu^{\mu'}_x-\nu^\mu_x}{\rho_x} \right) |n-\nabla_{\mu', \mu}\rangle\langle n| \nonumber.
\end{align}
We therefore have that $A_1$, $A_2$, $B$, as well as their commutators, are of order $V$, as would be any higher order commutator. We can then apply the Baker-Campbell-Hausdorff formula to second order in $\delta t$ to sums of those operators, such as $A_1+A_2$:
\begin{equation}
\mathrm{e}^{\delta t(A_1+A_2)}\sim\mathrm{e}^{\delta tA_1}\mathrm{e}^{\delta tA_2}\mathrm{e}^{-\delta t^2[A_1,A_2]/2}\sim\mathrm{e}^{\delta tA_2}\mathrm{e}^{\delta tA_1}\mathrm{e}^{\delta t^2[A_1,A_2]/2}.
\end{equation}
In both expressions, the last term is negligible for $\delta t\rightarrow 0$, which concludes the proof. A similar computation leads to the commutation of $A^{\gamma'\gamma}_{\mathfrak{s}}$ with $B_h$.

~

Let us now estimate the value of a term of the form $\langle n_2|\mathrm{e}^{\delta t(A^{\gamma'\gamma}_{\mathfrak{s}}+B^{\gamma'\gamma})}|n_1\rangle$ in the limits stated above, also considering that $V\delta t\rightarrow\infty$, which is to say that the number of reaction occurring in one time-step is large. This is an important step to ensure that we are in a large deviations regime. Applying $B$ to $n_1$ and expanding the other exponential, we find
\begin{align}
\langle n_2|\mathrm{e}^{\delta t(A^{\gamma'\gamma}_{\mathfrak{s}}+B^{\gamma'\gamma})}|n_1\rangle&\asymp\mathrm{e}^{-V\delta t k_{\gamma'\gamma}\rho_1^{\nu^\gamma}}\sum_k\frac{\delta t^k}{k!}\langle n_2|( A^{\gamma'\gamma}_{\mathfrak{s}})^k|n_1\rangle \nonumber\\
&\asymp\mathrm{e}^{-V\delta t k_{\gamma'\gamma}\rho_1^{\nu^\gamma}}\frac{\left(V\delta t~\mathrm{e}^{\mathfrak{s}_{\gamma' \gamma}}k_{\gamma'\gamma}\rho_1^{\nu^\gamma}\right)^k}{k!}~\mathbb{I}\bigl(n_2-n_1-k\nabla_{\gamma', \gamma}\bigr).
\end{align}
Given that, for a large factor $X$, the function $X^k/k!$ has a maximum around $k=X$, we can rescale $k$ by $X=V\delta t$ in the expression above, defining
\begin{equation}
k=V\delta t~\lambda_{\gamma', \gamma}
\end{equation}
where, as before, $\lambda$ has the interpretation of a number of transitions per time-step, rescaled by volume. One final use of the Stirling formula then yields
\begin{equation}
\langle n_2|\mathrm{e}^{\delta t(A^{\gamma'\gamma}_{\mathfrak{s}}+B^{\gamma'\gamma})}|n_1\rangle\asymp\exp\Bigl(-V\delta t ~ \tilde{L}_\mathfrak{s}(\lambda_{\gamma', \gamma},\rho_1) \Bigr)~\delta\bigl(\rho_2-\rho_1-\delta t~\lambda_{\gamma', \gamma}\nabla_{\gamma', \gamma}\bigr).
\end{equation}
where we have replaced the discrete indicator function $\mathbb{I}$ with a continuous delta function, and
\begin{equation}
\tilde{L}_\mathfrak{s}(\lambda_{\gamma'\gamma},\rho)=\lambda_{\gamma'\gamma}\ln(\lambda_{\gamma'\gamma}/k_{\gamma'\gamma}\rho^{\nu^\gamma})- \lambda_{\gamma'\gamma}+k_{\gamma'\gamma}\rho^{\nu^\gamma}-\mathfrak{s}_{\gamma' \gamma}\lambda_{\gamma'\gamma}.
\end{equation}
This is the rate function of a single biased Poisson distribution, which is what we would expect for independent jumps processes.
Similarly, we have
\begin{equation}
\langle n_2|\mathrm{e}^{\delta t(A^{\gamma'\gamma}_{\mathfrak{s}}+B^{\gamma'\gamma})}|n_1\rangle\asymp\mathrm{e}^{V\delta t ~h\rho_1}~\delta\bigl(\rho_2-\rho_1\bigr).
\end{equation}

We can then consider the whole of $\langle n_f|\mathrm{e}^{\delta tW_{\mathfrak{s},h}}|n_i\rangle$, by splitting the terms in $W_{\mathfrak{s},h}$ in any order we like and injecting the identity between every term. We are left with the product of all exponentials, with a global delta function that takes into account all the currents $\lambda_{\gamma'\gamma}$:
\begin{equation}
\langle n_f|\mathrm{e}^{\delta tW_{\mathfrak{s},h}}|n_i\rangle\asymp\exp\Bigl(-V\delta t ~ L_\mathfrak{s}(\lambda,\rho_i) \Bigr)~\delta\bigl(\rho_f-\rho_i+\delta t~\nabla\cdot\lambda\bigr).
\end{equation}
with
\begin{equation}
L_\mathfrak{s}(\lambda,\rho)=\sum\limits_{\gamma',\gamma}\lambda_{\gamma'\gamma}\ln(\lambda_{\gamma'\gamma}/k_{\gamma'\gamma}\rho^{\nu^\gamma})- \lambda_{\gamma'\gamma}+k_{\gamma'\gamma}\rho^{\nu^\gamma}-\mathfrak{s}_{\gamma' \gamma}\lambda_{\gamma'\gamma}-\sum_x h_x\rho_x.
\end{equation}

All that remains is to define $\rho_f-\rho_i=\dot\rho~\delta t$ to recover formulae \eqref{detLag} for $\mathfrak{s}=0$ and $h=0$, and \eqref{detBiasLag} otherwise.

~

To obtain an expression for the Hamiltonians instead, one may simply do a Legendre transform of the Lagrangians with respect to their respective flux variables. However, we may also start from their definition as a generating function:
\begin{equation}
H_\mathfrak{s}(\sigma,n_i)=\frac{1}{V\delta t}\ln\Bigl(\sum_n\langle n|\mathrm{e}^{\delta tW_{\sigma+\mathfrak{s},h}}|n_i\rangle\Bigr).
\end{equation}
In this case, the sum over the endpoint $n$ means we do not have to keep track of the number of particles exchanged. The same procedure as above, summed over the endpoint, yields
\begin{align}
\sum_n\langle n|\mathrm{e}^{\delta t(A^{\gamma'\gamma}_{\sigma+\mathfrak{s}}+B^{\gamma'\gamma})}|n_1\rangle&\asymp\mathrm{e}^{-V\delta t k_{\gamma'\gamma}\rho_1^{\nu^\gamma}}\sum_{k,n}\frac{\delta t^k}{k!}\langle n|( A^{\gamma'\gamma}_{\mathfrak{s}})^k|n_1\rangle \nonumber\\
&\asymp\mathrm{e}^{-V\delta t k_{\gamma'\gamma}\rho_1^{\nu^\gamma}}\sum_k\frac{\left(V\delta t~\mathrm{e}^{\sigma_{\gamma' \gamma}+\mathfrak{s}_{\gamma' \gamma}}k_{\gamma'\gamma}\rho_1^{\nu^\gamma}\right)^k}{k!}\nonumber\\
&\asymp\exp\Bigl(V\delta t~ k_{\gamma'\gamma}\rho_1^{\nu^\gamma}(\mathrm{e}^{\sigma_{\gamma' \gamma}+\mathfrak{s}_{\gamma' \gamma}}-1) \Bigr). \nonumber
\end{align}
Putting all terms together including $B_h$ immediately yields
\begin{equation}
H_\mathfrak{s}(\sigma,n)=\sum\limits_{\gamma',\gamma}k_{\gamma'\gamma}\rho_1^{\nu^\gamma}(\mathrm{e}^{\sigma_{\gamma' \gamma}+\mathfrak{s}_{\gamma' \gamma}}-1)+\sum_x h_x\rho_x
\end{equation}
which is eq.\eqref{detBiasHam}. Setting $\mathfrak{s}=0$ and $h=0$ yields eq.\eqref{detHam}. Finally, setting $\sigma=\nabla f$ yields eq.\eqref{Ham}.

~~

One should note that these derivations very much rely on the fact that $n$ stays of order $V$ at all times, so that a change of $n_x$ of order $\nu_x^\gamma$ is negligible in terms of $\rho$. For dynamics whose trajectories end up close to any wall $n_x=0$, one has to treat that species separately at the microscopic scale, resulting in a so-called \textit{hybrid dynamics} \cite{Bressloff2014}.

It should also be noted that those results can be easily generalised to dynamics other than mass-action, as long as one is able to show that all commutators between reaction operators are of order $V$ (or of the same order as the operator themselves).

\bibliographystyle{mybibstyle}

\bibliography{Biblio}{}

\end{document}